\documentclass[fleqn,10pt]{article}
\usepackage[utf8]{inputenc}
\usepackage[T1]{fontenc}
\usepackage{authblk}
\usepackage{chngcntr}
\usepackage[toc,page]{appendix}

\usepackage{array}
\usepackage{arydshln}
\usepackage{float}
\usepackage[]{graphicx}
\usepackage{amsmath}
\usepackage{amssymb}
\usepackage{verbatim}
\usepackage[sort&compress]{cleveref}
\usepackage{bbm, bm}
\usepackage{algpseudocode, algorithm}
\usepackage{booktabs}
\usepackage{multirow}
\usepackage{tabularx}
\usepackage[export]{adjustbox}
\usepackage{appendix}

\usepackage{geometry}
\usepackage[caption=false]{subfig}
\usepackage{color}
\usepackage{url}

\graphicspath{{./images/}}

\Crefname{appsec}{appendix}{appendices}

\newcommand{\R}{\mathbb{R}}
\newcommand{\cF}{\mathcal{F}}
\newcommand{\cL}{\mathcal{L}}
\newcommand{\cN}{\mathcal{N}}

\newcommand{\der}{\mathrm{d}}

\newcommand{\Ytraj}{Y_{1:T}}

\newcommand{\given}{\,|\,}
\newcommand\iid{i.i.d.~}

\numberwithin{equation}{section}

\title{Inferring the spread of COVID-19:
the role of time-varying reporting rate
in epidemiological modelling}
\author[1,*]{Adam Spannaus}
\author[2,1,3]{Theodore Papamarkou}
\author[4]{Samantha Erwin}
\author[1]{J. Blair Christian}

\affil[1]{Computational Sciences and Engineering Division, Oak Ridge National 
	Laboratory, Oak Ridge, TN, USA}

\affil[2]{Department of Mathematics, The University of Manchester, Manchester,
	UK}
	
\affil[3]{Department of Mathematics, University of Tennessee, Knoxville, TN, USA}

\affil[4]{Pacific Northwest National Laboratory, Richland, WA, USA}

\begin{document}
	\maketitle	
\begin{abstract}

The role of epidemiological models is crucial for 
informing public health officials during a public health emergency,
such as the COVID-19 pandemic. However, traditional epidemiological models fail to capture the time-varying effects of mitigation strategies and do not account for under-reporting of active cases, thus introducing bias in the estimation of model parameters. 
To infer more accurate parameter estimates and to reduce the uncertainty of these estimates, 
 we extend the SIR and SEIR epidemiological models with two time-varying parameters that capture the transmission rate and the rate at which active cases are reported to health officials.
Using two real data sets of COVID-19 cases,
we perform Bayesian inference via our SIR and SEIR models with time-varying transmission and reporting rates and
via their standard counterparts with constant rates;
our approach provides parameter estimates with more realistic interpretation,
and one-week ahead predictions with reduced uncertainty. Furthermore, we find consistent
under-reporting in the number of active cases in the data that we consider,
suggesting that the initial phase of the pandemic was more widespread than previously reported.

\end{abstract}

\section{Introduction}


During a disease outbreak, epidemiological forecasting informs public health decisions by 
predicting how widely the disease will possibly spread~\cite{astolfi2012informing}. Such 
information is critical for public health officials trying to understand the dynamics through 
which a disease is transmitted and to slow its propagation~\cite{anderson1992infectious}. 
It is also important to quantify the proportion of infections in a community and the 
frequency with which infections are being reported to public health officials.
Moreover, in the case of
a novel disease,
there are uncertainties
in estimates of the infection rate and
of the incubation period,
and reliable testing strategies have yet to 
be developed. 
These considerations 
are relevant during the  
SARS-CoV-2 (COVID-19) pandemic as the uncertainty surrounding the disease's transmission and incubation 
rates have made coordinating the global response to this disease a substantial challenge to
public health officials.

One of the most widely 
employed mathematical models of general disease transmission is the Susceptible-Infected-Removed 
(SIR) model~\cite{kermack1927contribution}. This model describes the interactions between three population groups: those who are are susceptible to an infection, those who are infected, and those who have been removed from the population through either recovery or death. Over time, the SIR model has been augmented to include other groups,
such as an exposed individuals,
asymptomatic carriers, or
individuals immune to the infection~\cite{blackwood2018introduction}.
In the present context of the COVID-19 pandemic, 
a latency period in the course of the disease has been
proposed~\cite{rothe2020transmission},
during which an individual carrying the disease does not present symptoms.
This suggests the possibility of including an `exposed' state,
thus modelling the spread of COVID-19 with a
Susceptible-Exposed-Infected-Resistant (SEIR) model.

Deterministic epidemiological models cannot reflect the real-time implications of 
preventative measures, such as safer-at-home orders, social distancing, or mask mandates that
directly impact the transmission rate~\cite{andersson2012stochastic}.
Moreover, these models do not take into account 
climatic changes, nor social cycles, e.g., holiday gatherings or the periodic starting and stopping of the school calendar.
To capture the time-varying effects of preventative measures and changes in social interactions, 
we choose stochastic state space representations of both the SIR and SEIR model and 
adopt a Bayesian approach.

The novelty in our model is that the rate of transmission and
the reporting rate of positive cases vary in time.
Time-varying transmission and reporting rates, as opposed to constant ones,
provide more flexible modelling assumptions, and
thus adhere more closely to respective observed rates in COVID-19 data.
By allowing the transmission rate to vary in time,
our model adapts to
heterogeneity of disease outcomes,
to mutations in the disease,
to asymptomatic transmission,
and to mitigation strategies used to slow the spread of the disease.
However, a time-varying transmission rate does not fully capture the spread of
a disease though a population.
Indeed, not all cases are reported due to faulty tests, 
inability to obtain a test, or false negative results for example.
In order to further quantify disease proliferation
through a susceptible population, 
we also allow the reporting rate to vary in time.
By including a time-varying reporting rate,
our model is more flexible to adapt to
under-reporting of cases and
to advancements in testing reliability.
In other words,
this time-varying reporting rate allows our model to dynamically adapt to changes in the
spread of COVID-19,
thus informing more accurately
the reporting of positive cases to the health agencies.

During the COVID-19 pandemic, and specifically in the early phases,
a lag in testing times and under-reported case rates
have been observed. Similar studies on the Zika virus note
a high occurrence of under-reporting and
estimate reporting rates by including
a separate unreported infected population
in epidemiological modelling~\cite{shutt2017}.
Similarly, in estimating the reporting rate for 
shigellosis, a secondary unreported infected 
compartment is proposed by~\cite{joh2013}. An
alternative approach is taken in a study 
of an influenza pandemic, where a functional form of the reporting 
rate is estimated as either linearly increasing or constant 
depending on the time-frame of the pandemic~\cite{chong2014,saberi2020accounting}.
Rather than introducing
an additional compartment into an epidemiological model,
we instead employ a time-varying parameter of reporting rate.
So, we avoid a structural change to our model,
while making the model more flexible via time-varying parameterization.
Augmenting an epidemiological model is not the only approach to estimate
the spread of a disease throughout a population. Alternatively, one could develop
an artificial neural network~\cite{sabir2022artificial}, or a 
fractional-order epidemiological model~\cite{ali2022investigation}.




To demonstrate our approach,
we study two regional outbreaks with different mitigation strategies
using the COVID-19 case counts, as collated by the New York Times~\cite{nytimes-covid}.
Considering data from Tennessee and New York showcases the flexibility of our Bayesian model with time-varying transmission and reporting rates,
because the spread of COVID-19 in these states presents different modelling challenges.
Notably, these two states experienced the initial six months of the pandemic
in widely differing fashion.
Indeed, each state handled the beginning of the pandemic differently, 
employed different testing strategies, and continue to have different reopening policies~\cite{raifman2020covid}.
Such widely varying pandemic management strategies are demonstrated by the 
COVID-19 data sets associated with Tennessee and New York, which
exhibit drastically different dynamics of case counts.
Our experiments highlight the capacity of our Bayesian state space epidemiological models
with time-varying transmission and reporting rates
to fit data representing different dynamics of disease spread,
to estimate the transmission rate and the reporting rate
during the progression of COVID-19,
and to reduce the uncertainty in predictions of COVID-19 cases.



\section{Model formulation}\label{sect:model_form}

\subsection{S(E)IR epidemiological model}\label{sect:epi-model}


To investigate the spread of COVID-19 through a population, we
use the traditional SIR and SEIR epidemiological models, augmenting them with two temporal variables.
One of these variables accounts for the rate of
transmission,
while the other variable
quantifies the rate at which positive cases are 
reported.
These time-varying variables account
for variation in the transmissibility of a disease and
in the rate at which
new infections are reported. 
Consequently, our model is 
able to account for variation and uncertainty present during
a public health emergency.


Allowing the transmission rate of COVID-19 to vary in time,
similar to the approach of~\cite{del2015sequential,dureau2013capturing,funk2018real,mishra2021comparing},
we are able to capture intervention
measures enacted by public health officials, such as mask mandates or 
shelter-in-place orders, or `super-spreader' events which have direct impacts
on disease prevalence.
By modelling the rate at which the disease spreads as a time-varying variable, 
we may better quantify the spread of the disease, which in turn, yields more
accurate information about the course of the pandemic to public health officials concerned with slowing community spread.

Quantifying the transmissibility of a disease does not fully capture its
reach and spread, though. If we consider the early phases of the 
COVID-19 pandemic, wide-spread access to testing was unavailable, and
disease prevalence within communities was widely under-reported~\cite{fisher2020global,lau2020evaluating}.
To this end, we present a novel S(E)IR epidemiological model
by introducing two time-varying variables, namely
the \textit{reporting rate} $p(t)$ and
the \textit{transmission rate} $\beta (t)$.
The reporting rate $p(t)$
captures the percentage of positive cases reported,
and the transmission rate $\beta (t)$
quantifies the rate at which COVID-19 spreads through a population 
at time $t$.





The SIR epidemiological model with time-varying transmission rate
$\beta (t)$
is represented by the
system of differential equations
\begin{equation}
	\begin{aligned}\label{eq:SIR}
		\der S(t) &= -\beta(t)S(t)\frac{I(t)}{N} \der t, \\
		\der I(t) &= (\beta(t)S(t)\frac{I(t)}{N} - \gamma I(t)) \der t,\\
		\der R(t) &= \gamma I(t)\, \der t,\\
		\der w(t) &= \mu(w(t), \theta_w) \der t +
		\sigma(w(t),\theta_w)\,\der B(t), \quad
		w(t) = g(\beta(t)),
	\end{aligned}
\end{equation}
where $w(t)$ is a stochastic differential equation (SDE),
which entails the Wiener process $B(t)$ and
controls the transmission rate~\cite{oksendal2013stochastic}. 
In this SDE, we define drift $\mu$ and diffusion $\sigma$ terms,
parameterized by $\theta_w$, and a function $g:\R^+\to\R$. 
It is this time-varying parameter
that controls 
the extent to which a disease spreads throughout a population.
Lastly, to fully specify the epidemiological model, we write $S(t)$ as individuals
susceptible to COVID-19, $I(t)$ as individuals infected with COVID-19, 
and $R(t)$ as individuals removed from the population who are no longer able to become infected. 
Susceptible individuals move into the infected compartment at 
rate $\beta (t)$
and infected individuals become removed at rate $\gamma$.

An expansion on the SIR model is the SEIR model which
includes an additional `exposed' compartment $E(t)$
between the `susceptible' and `infected' compartments.
Individuals who are in the disease's latent period, the `exposed' compartment,
move into the `infected' compartment with rate $\kappa$ and are and thus are capable of infecting other susceptible individuals.
The system of differential equations that represents the SEIR model can be found
in appendix~\ref{seir_model_def}.
The salient difference between the SIR and SEIR models is the inclusion of the `exposed' state in the latter;
individuals move into this compartment
once they have been exposed to an infected individual,
but are not yet expressing any symptoms.  
Such a latent phase is pertinent when quantifying the breadth of the 
pandemic due to the challenges in correctly accounting for such
individuals. Moreover, the `exposed' compartment not only acts as a delay between 
the `susceptible' and `infected' compartments, but individuals in this
compartment may transmit the disease before becoming actively infected themselves.

In this study, for both SIR and SEIR models, 
we take $\mu=0$ and
$\sigma(w(t),\theta_w)=\theta_w$,
thus assuming that the transmission rate 
on any day is likely to be the same as the previous day, and set $g(\cdot) = \log(\cdot)$.
By choosing $\mu=0$ and
$\sigma(w(t),\theta_w)=\theta_w$,
the resulting path 
$w(t)$, defined by the SDE
$\der w(t)=\theta_w\der B(t)$, is a Brownian motion.
Allowing $w(t)$ to vary in time defines
the effective transmission rate $\beta(t)=g^{-1}(w(t))$,
which controls the extent to which
a disease spreads between individuals within a population.

\subsection{State space model}\label{sect:SS_model}




Incorporating temporal information about the dynamics driving the spread 
of COVID-19 into our model, we discretize the path of $w(t)$ 
which defines the transmissibility, and seek to infer $p(t)$ 
at each time step. 
These two correlated time series
quantify the rate at which COVID-19 spreads through
a population~\cite{gostic2020practical,saberi2020accounting}. 
Then coupling these time series with observed case counts as
reported by public health agencies, we adopt a state-space modelling paradigm
for our inference problem~\cite{birrell2018evidence}.

A state space model relates two discrete time processes by a probabilistic model
incorporating both state evolution and observation densities.
In the present context, we are given the number of cases reported
by the public health officials, and seek to infer the
distribution of 
the reporting rate $p(t)$ and of the transmission rate $\beta (t)$.
We view
the transmission rate and reporting rate
as discrete time processes and write $\beta_t$ and $p_t$ for the discrete time 
counterparts of $\beta(t)$ and 
$p(t)$, respectively.
Making some regularity 
assumptions~\cite{zhigljavsky2007stochastic} on the evolution dynamics of the 
system given in equation~\eqref{eq:SIR}, and
denoting any sequence as
$\{c_t\}_{t\geq 0}$ for $i\leq j$ as  $c_{i:j}= (c_i, c_{i+1},\dots, c_j)$,
we view the stochastic epidemiological model as a state space model, written
\begin{equation}\label{eq:seir-hmm}
	\begin{aligned}
		\der w_t &= \theta_w\,\der B_t,\ \qquad &\der p_t = p_t \der t + \vartheta^2\der W_t,\\ 
		Y_{1:T} &\sim h(Y_{1:T}\given X_{0:T}, p_{0:T}, \theta_{Y}), \qquad
		&X_{0:T} = \cF(w_{0:T}; \theta_w).
	\end{aligned}
\end{equation}
The transmissibility $w_t$ of COVID-19 is controlled by the SDE $\der w_t = \theta_w\,\der B_t$,
with Wiener process $B_t$, parameterized by $\theta_w$.
The reporting rate $p_t$ evolves according to the SDE $\der p_t = p_t\der t + \vartheta^2\der W_t$,
where $\vartheta$ is the diffusion parameter and $W_t$ is a Wiener process.
The reporting rate $p_t$, coupled with the number of observed cases $Y_{t}$,
and associated parameters $\theta_{Y_t}$ defines an
observation density, $Y_{t} \sim h (\cdot\given X_{t}, p_t, \theta_Y)$. 
We define $X_t$ as the latent number of cases, i.e., 
$X_t = \int_{t-1}^{t}\,\beta_{\tau}S_{\tau}\frac{I_{\tau}}{N}\der \tau$
or $X_t =\int_{t-1}^{t}\, \kappa E_{\tau} \der \tau, 1\leq t\leq T$
for an SIR or SEIR model, respectively, and $X_0\sim \pi_0$
for a prior density $\pi_0$.
We write this recursion as 
$\mathcal{F}(w_t;\theta_w)$ to make explicit that the solution to~\cref{eq:SIR}
depends on the state $w_t$ and can be computed for any value thereof via numerical
integration.

To define the observation density $h(\cdot\given X_t, p_t, \theta_Y)$, we first assume that  
the reporting of new cases are independent Bernoulli random variables, i.e., 
each case is reported with probability $p_t$.
Then the waiting time until the first reported case
is geometrically distributed with the same parameter $p_t$ and 
we are interested in the weekly incidence rate, conditional on
the number of reported cases $Y_t$. 
Since $Y_t$ is the sum of \iid geometrically
distributed random variables with parameter $p_t$, it then follows that 
$Y_t\given X_t\sim\mathrm{NegBin}(p_t, X_t)$.
Invoking the central limit theorem,
the observations $Y_t$ are approximately 
Gaussian with mean $p_t X_t$ and variance $p_t(1-p_t)X_t + (p_t X_t \eta)^2$.
In this scenario, the variance term contains an additional parameter
$\eta$, which describes
the over-dispersion within a population.
See the Appendix for a more detailed discussion about the distribution of $Y_t$.
Lastly, we define parameters $\theta_{Y} = (\eta, \vartheta)$ and
$\theta_X = (\kappa, \gamma, X_0)$, or $\theta_X = (\gamma, X_0)$,
for the SEIR or SIR model, respectively.

\subsection{Model parameters}\label{sect:model_params}


We now describe the parameters utilized in our model. 
Firstly, the vector $\theta_Y = (\eta, \vartheta)$
contains parameters utilized in our model
when investigating a temporally-varying reporting rate $p_t$. Here
$\eta$ is incorporated in the observation variance, specifying
over-dispersion within the observations, and
indicating heterogeneity within a population~\cite{bolker2008ecological,Breslow84}.
It is a common occurrence in count 
data~\cite{bolker2008ecological,Breslow84} and accounts for
large variances in individual outcomes. Furthermore it 
can signal the presence of `super-spreading' events~\cite{lloyd2007maximum}. 
Secondly, $\vartheta$ is the standard deviation of 
the reporting rate. By investigating the marginal density
of $p_t$ at each time step, we may quantify the uncertainty in the rate of reporting
of positive cases. 

We compare our model against a model with constant reporting rate,
denoted $p_c$,
for the entire time duration. In the model with constant reporting rate,
we set $\theta_Y = \eta$, and we may infer the variance \emph{a posteriori}
from Markov chain Monte Carlo (MCMC) integration, but this is 
a static value that is not capable of adapting to changes in the
realities of a pandemic, such as more sensitive or accurate testing methods and
the availability of individuals to obtain a test. 

For the parameters governing the movement between the compartments
of the epidemiological model, aside from the transmission rate $\beta_t$, we 
follow~\cite{moghadas2020implications} and 
assume that the distribution of $\gamma$, governing the movement of individuals
from the infected bin to recovered, follows a Gaussian distribution 
with mean $5.058$ days and standard deviation of $1.519$ days, with support
on the interval of $2.228$ and $11.800$ days. For our SEIR models, we 
sample the parameter $k$ from a gamma distribution with shape parameter
$1.058$ and scale $2.174$. The parameter $k$ controls the rate at which
individuals move from the disease's latent phase to an active infection.


\section{Results}\label{sect:numerics}


\subsection{Parameter uncertainty}

\begin{figure}
    \centering
    \subfloat[Weekly cases (data)]{\includegraphics[width=0.33\textwidth]{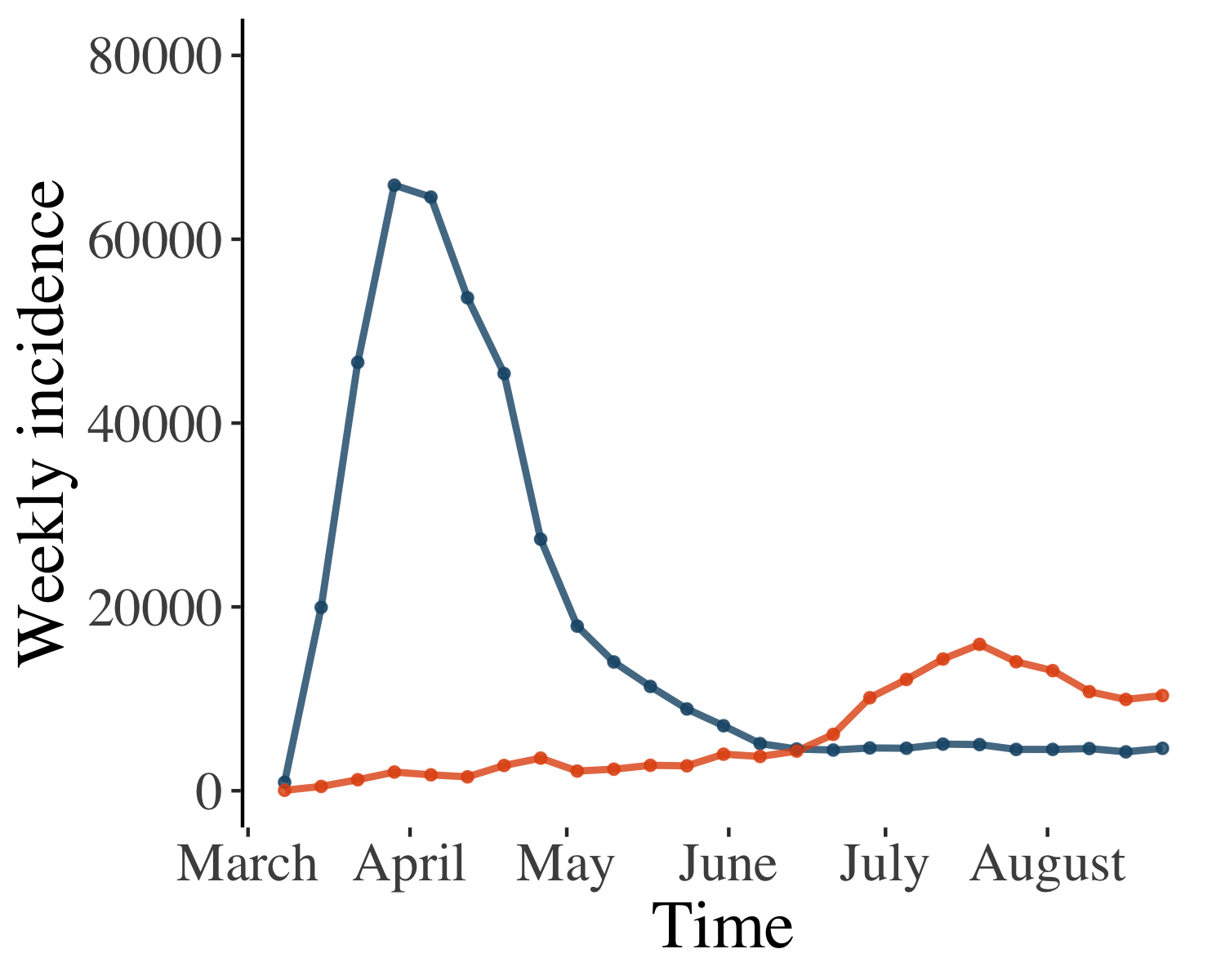}\label{fig:data}}%
    \subfloat[New York data reconstruction]{\includegraphics[width=0.33\textwidth]{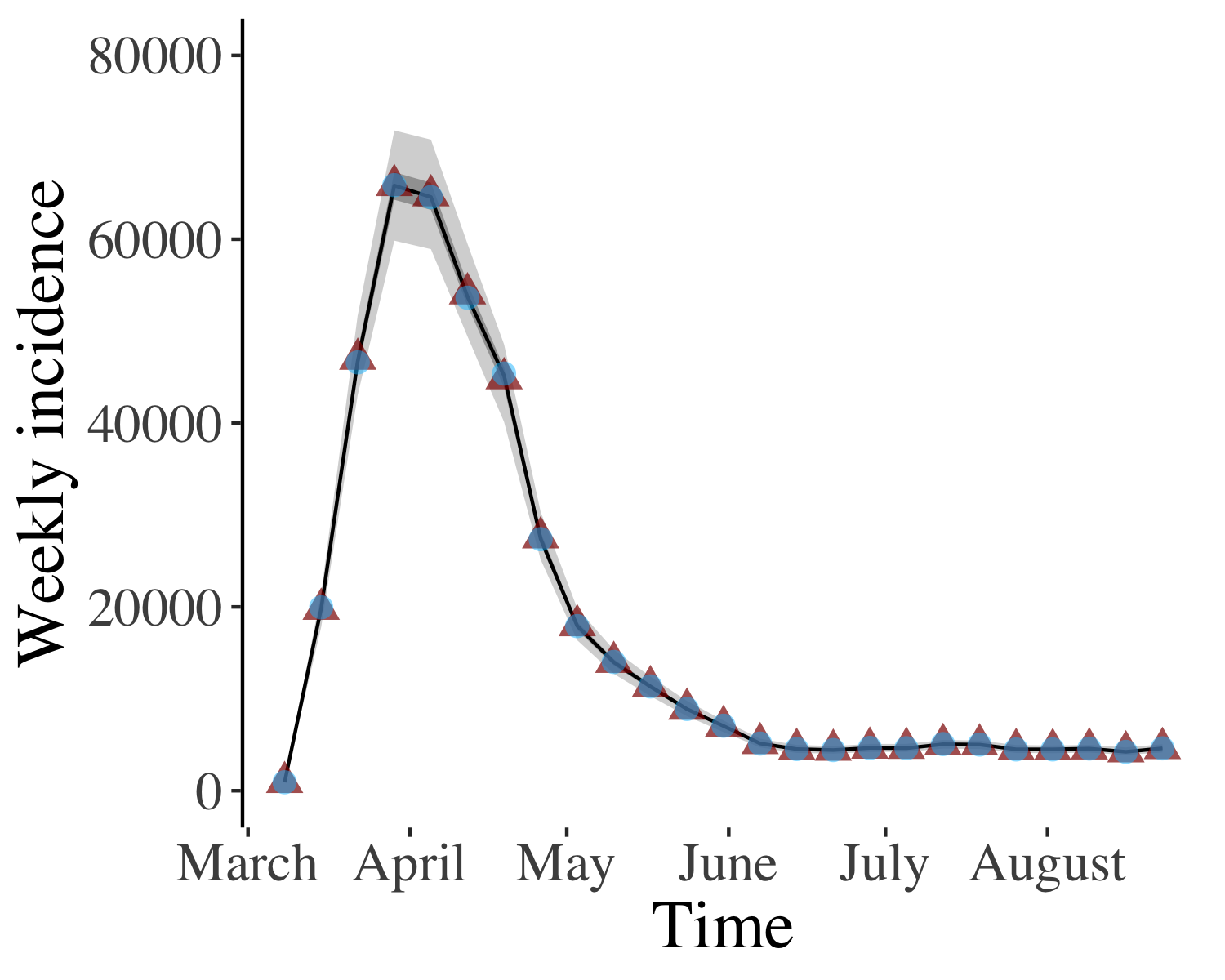}\label{fig:NY_obs}}%
    \subfloat[Tennessee data reconstruction]{\includegraphics[width=0.33\textwidth]{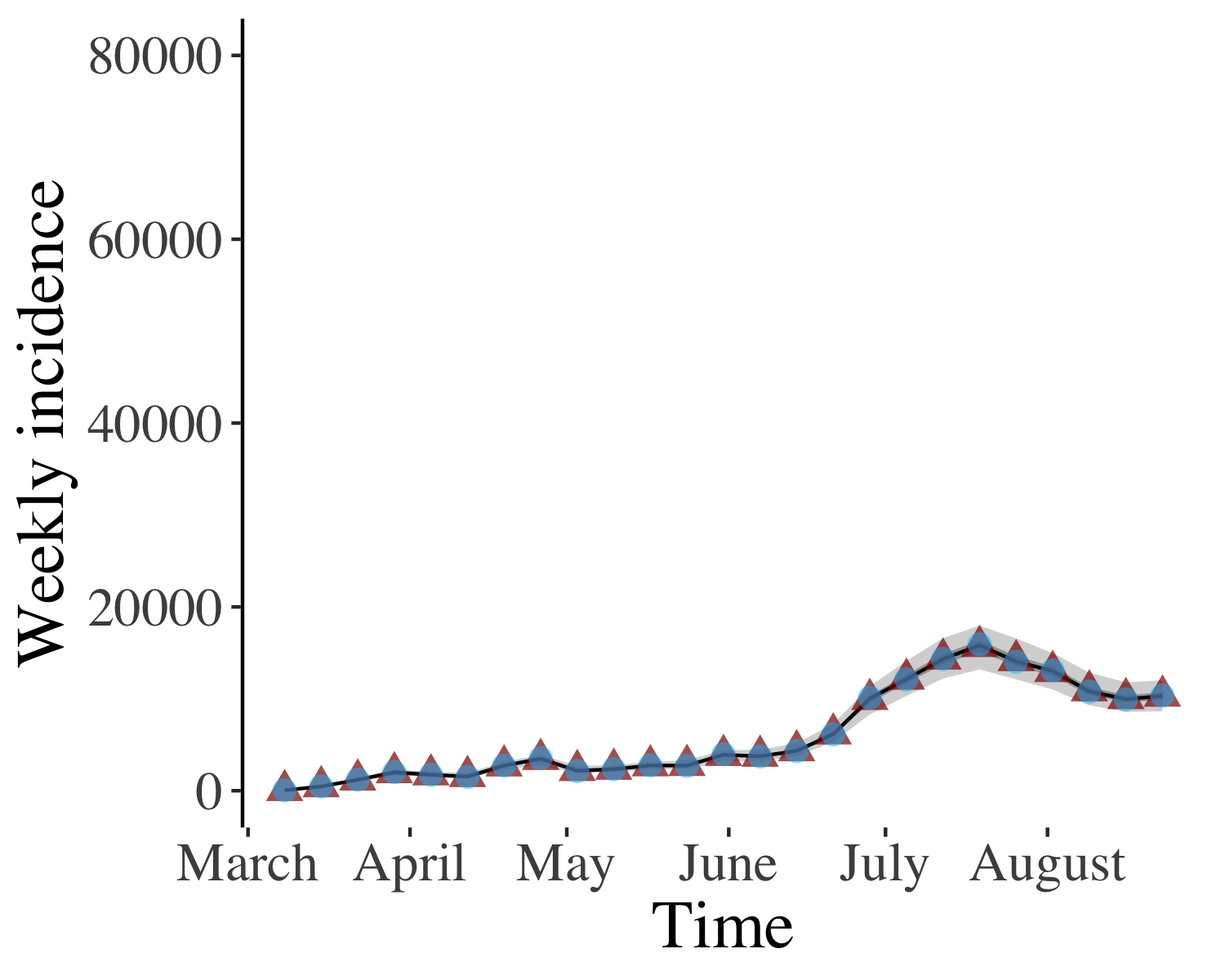}\label{fig:TN_obs}}%
    \caption{The weekly data of COVID-19 cases for New York (blue) and for Tennessee (orange)
    from March 1, 2020 through August 31, 2020 are shown in~\cref{fig:data}.
    The reconstructed weekly cases for New York and for Tennessee over the same time period,
    based on our SIR model with time-varying reporting rate $p_t$, are displayed in 
    ~\cref{fig:NY_obs} and~\cref{fig:TN_obs}, respectively.
    In~\cref{fig:NY_obs} and~\cref{fig:TN_obs}, which show our model-based data reconstruction,
    blue triangles,
    red triangles,
    black lines,
    light-shaded grey areas and dark-shaded grey areas
    represent the
    original data,
    posterior means,
    posterior medians,
    $75\%$ credible intervals
    and $95\%$ credible intervals, respectively.}
	\label{fig:ny_and_tn_case_cts}
\end{figure}

Two data sets of reported COVID-19 cases,
one from New York and one from Tennessee,
demonstrate different evolution dynamics. 
The pandemic went through a period
of sustained exponential growth in New York, primarily in the New York City metropolitan
region, then abated
to a near constant level in subsequent months. Furthermore during the initial wave in New York, testing was
not widely available, and the seven-day rolling average of positive test case peaked at nearly 50\% in
early April 2020~\cite{nystatepercentpos}. 
On the contrary, the Tennessee data,
which are representative of the evolution dynamics of case counts for the majority of other states, 
exhibit a slow initial increase, followed by a first wave in April and a
much larger increase in July.
The variations between the incidence data for these two states are visible in~\cref{fig:data}. These two datasets represent the initial phase of the pandemic
from two distinct perspectives.
Firstly, the dramatic increase in the NY case counts 
during March 2020 induces greater uncertainty in the parameters as compared with 
the gradual increase in the Tennessee case counts.
Secondly, the different levels of induced parameter uncertainty
enable us to study how such different levels of uncertainty propagate
in time through our model.

\begin{figure}
	\centering
	\subfloat[]
	{\label{fig:NY_beta_const}
		\includegraphics[width=0.46\textwidth]{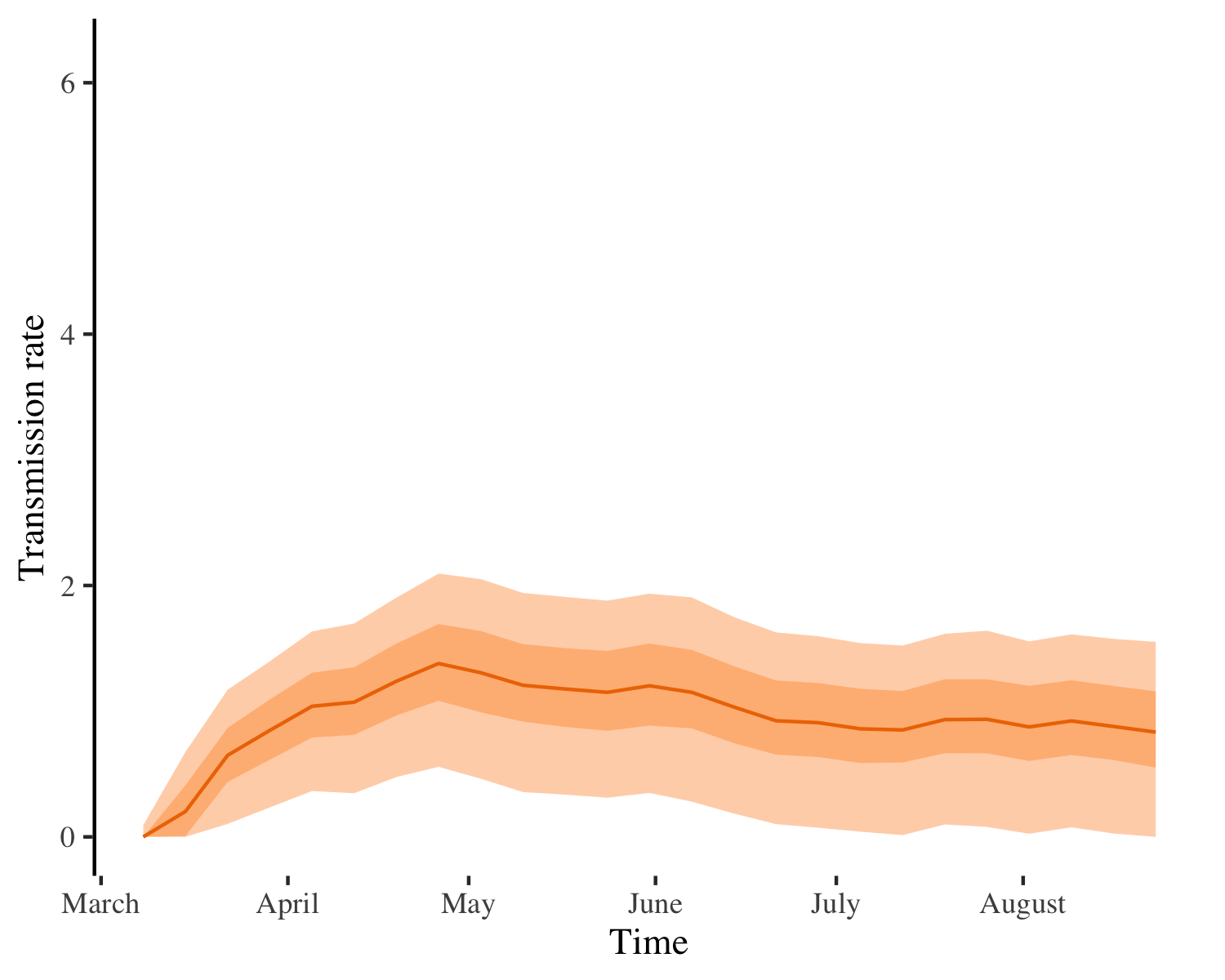}
	}
	\hfill
	\subfloat[]
	{\label{fig:NY_beta_SS}
		\includegraphics[width=0.46\textwidth]{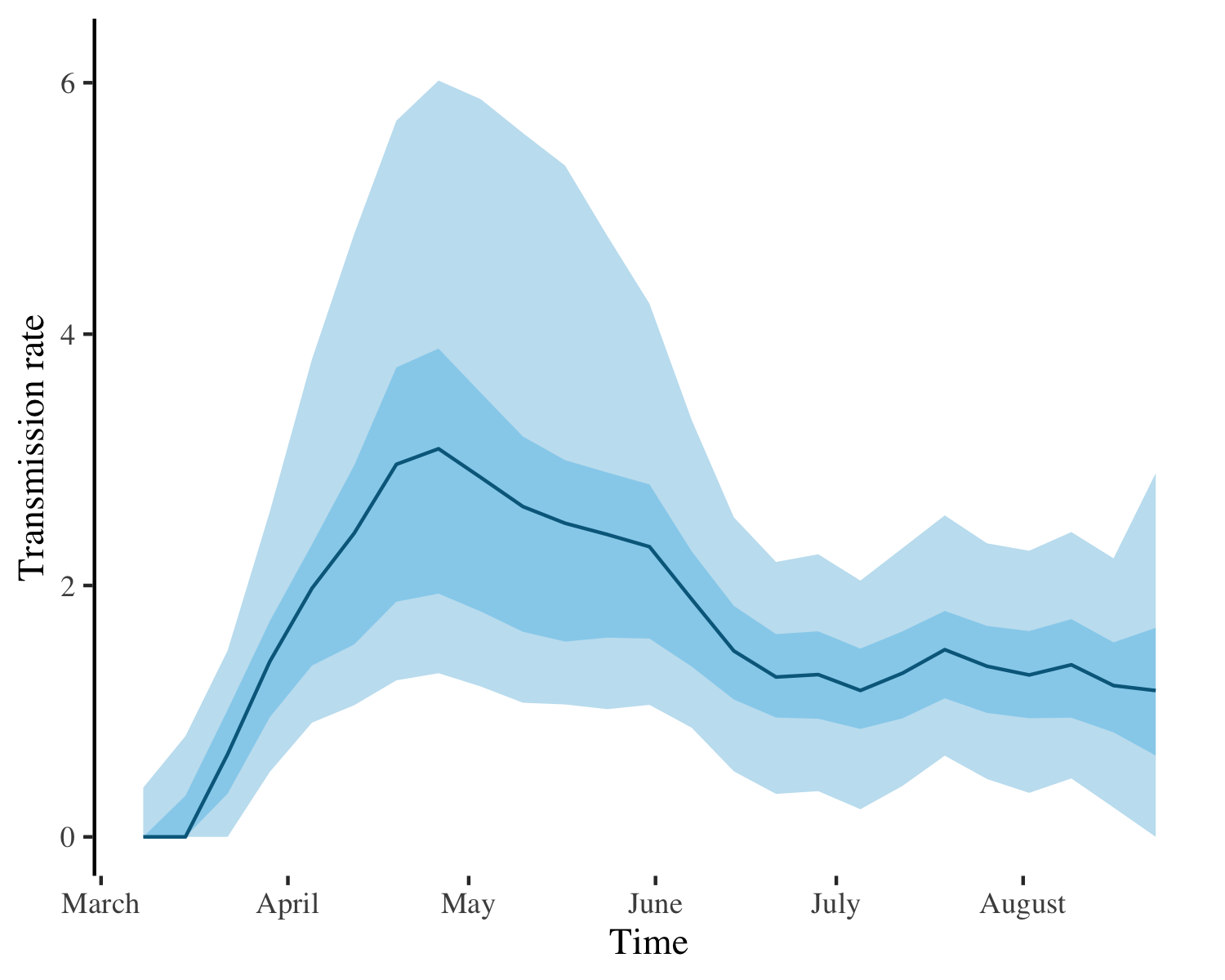}
	}
	\caption{Estimates of the COVID-19 transmission rate for New York
		based on the typical SIR model with constant reporting rate~(\cref{fig:NY_beta_const}) and
		based on our SIR model with a time-varying reporting rate $p_t$~(\cref{fig:NY_beta_SS}).
		Solid lines, light-shaded and dark-shaded areas
		correspond to posterior means, $75\%$ and $95\%$ credible intervals
		of the associated transmission rates.}\label{fig:NY_beta}
\end{figure}

The estimated transmission rate for New York
based on our SIR model with time-varying reporting rate $p_t$
exhibits a spike in transmissibility in April and May (\cref{fig:NY_beta_SS}),
which agrees with the spike of observed COVID-19 cases in April and May (blue line in \cref{{fig:data}}).
The typical SIR model with constant reporting rate
produces a transmission rate estimate that
fails to capture this spike in transmisibility, as seen in~\cref{fig:NY_beta_const}.
Recall that posterior estimates of the transmission rate quantify the rate
at which susceptible individuals move from the susceptible compartment to an 
active stage of infection.
As demonstrated by \cref{{fig:data}} and~\cref{fig:NY_beta},
our SIR model with time-varying reporting rate provides transmission rate estimates
that reflect the observed dynamics of transmission more faithfully than
a SIR model with constant reporting rate.

Furthermore, the higher rate of change in the number of observed COVID-19 cases in New York
between March and May
(blue line in \cref{{fig:data}})
induces higher uncertainty in the estimation of the transmission rate.
Our SIR model with time-varying reporting rate yields wider credible intervals
~(\cref{fig:NY_beta_SS})
for the estimated transmission rate in New York between March and May
in comparison to the SIR model with constant reporting rate
~(\cref{fig:NY_beta_const}).
Thus, letting the reporting rate vary with time facilitates the detection
of periods of higher uncertainty in transmission rate estimates.

\begin{figure}
	\centering
	\subfloat[]
	{\label{fig:TN_beta_const}
		\includegraphics[width=0.46\textwidth]{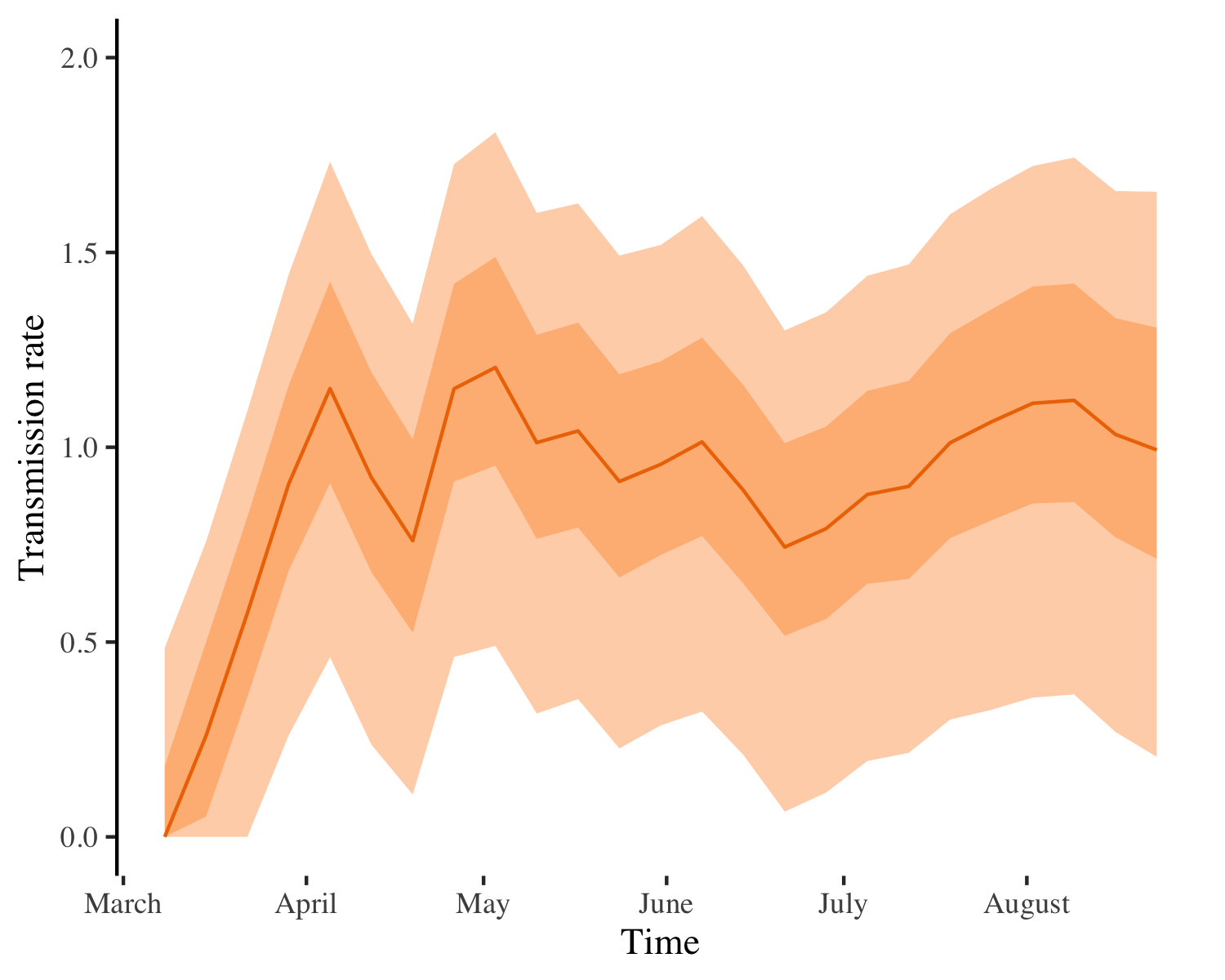}
	}
	\hfill
	\subfloat[]
	{\label{fig:TN_beta_SS}
		\includegraphics[width=0.46\textwidth]{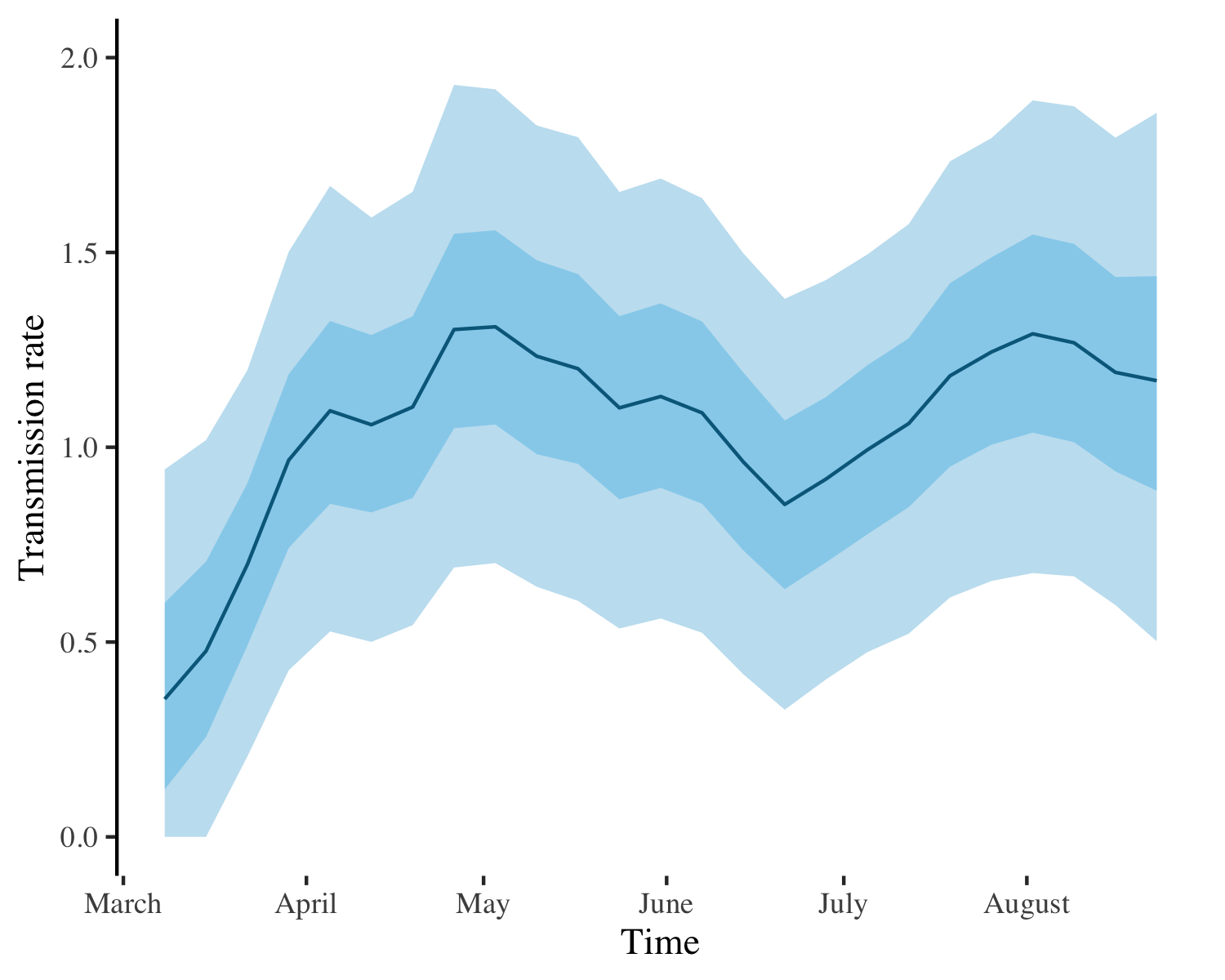}
	}
	\caption{Estimates of the COVID-19 transmission rate for Tennessee
		based on the typical SIR model with constant reporting rate~(\cref{fig:TN_beta_const}) and
		based on our SIR model with a time-varying reporting rate $p_t$~(\cref{fig:TN_beta_SS}).
		Solid lines, light-shaded and dark-shaded areas
		correspond to posterior means, $75\%$ and $95\%$ credible intervals
		of the associated transmission rates.}\label{fig:TN_beta}
\end{figure}

\begin{figure}
	\centering
	\subfloat[New York]
	{\label{fig:NY_SIR_rep}
		\includegraphics[width=0.46\textwidth]{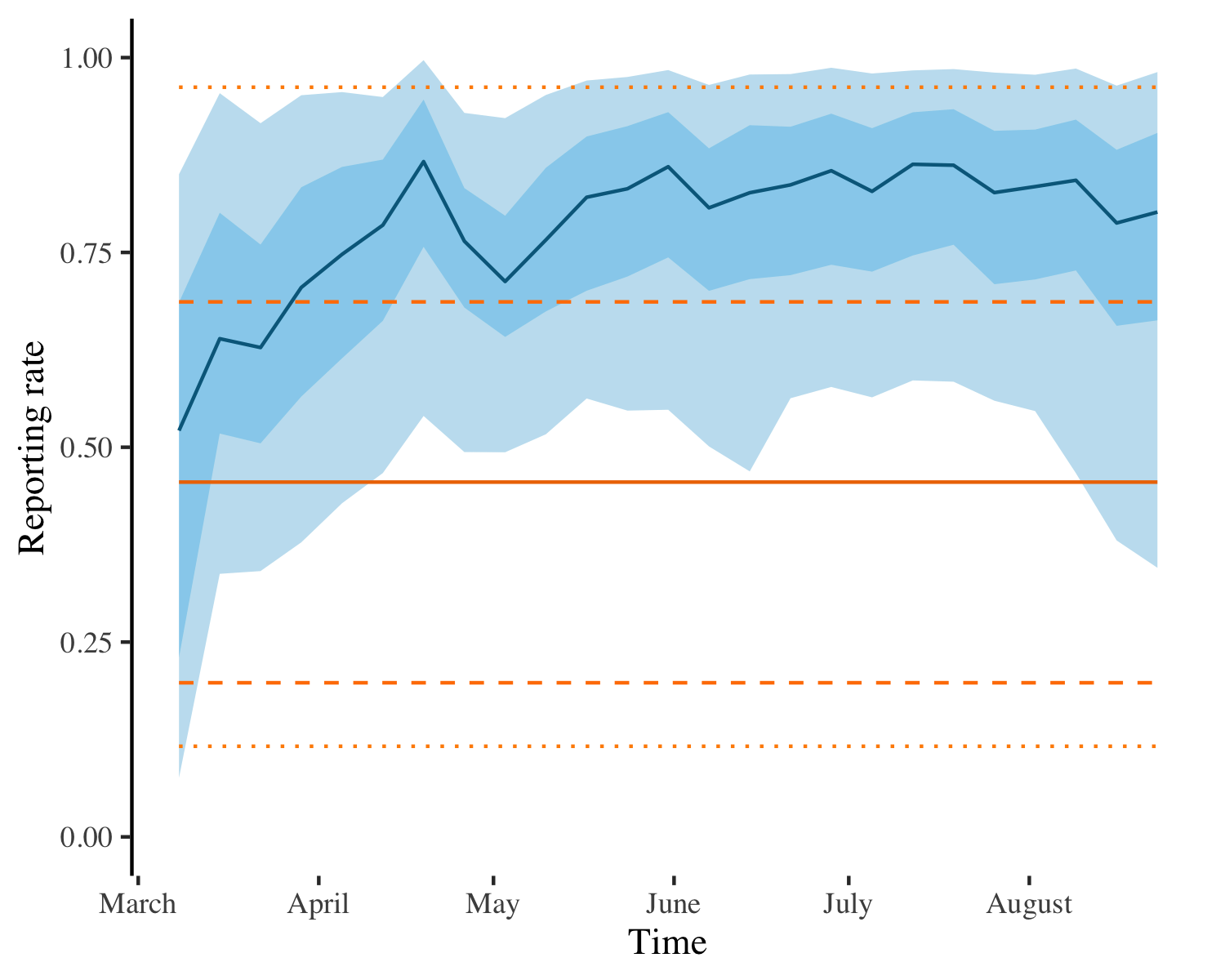}
		
	}
	\hfill%
	\subfloat[Tennessee]
	{\label{fig:TN_SIR_rep}
		\includegraphics[width=0.46\textwidth]{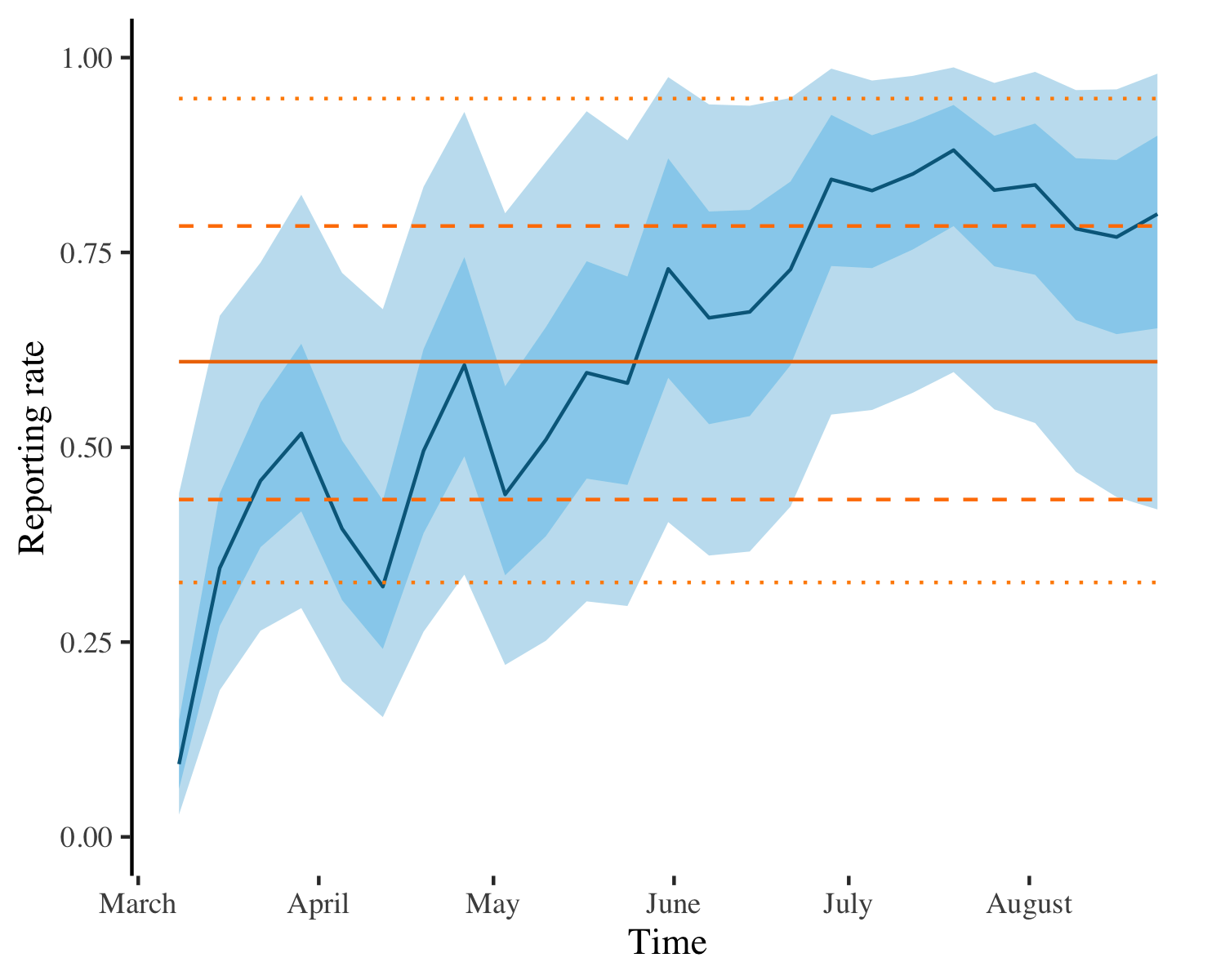}
	}
	\caption{Estimates of the COVID-19 reporting rate
	for New York~(\cref{fig:NY_SIR_rep}) and for Tennessee~(\cref{fig:TN_SIR_rep})
	obtained from the SIR model with constant (orange) and with time-varying (blue)
    reporting rate. 
	The blue line, light-shaded and dark-shaded blue areas
	represent the respective posterior mean,
	$75\%$ and $95\%$ credible intervals of the time-varying reporting rate.
	The orange solid, orange dashed and orange dotted line
	represent the respective posterior mean,
	$75\%$ and $95\%$ credible intervals of the constant reporting rate.
	}\label{fig:SIR_reporting}
\end{figure}

The transmission rate estimates obtained by 
fitting the SIR model with time-varying and with constant reporting rate 
to the Tennessee COVID-19 case data agree with one another
for the period between May and August
(see~\cref{fig:TN_beta_const} and~\cref{fig:TN_beta_SS}).
However, the SIR model with time-varying reporting rate
estimates a 
smaller drop in the transmission rate in April
as compared with the SIR model with constant reporting rate.
The former model exhibits a reduced reporting rate estimate
in April~(blue line in \cref{fig:TN_SIR_rep}).
So, both models capture the small drop in the number of COVID-19 cases
observed in Tennessee during April (orange line in~\cref{fig:data});
the flexible SIR model with time-varying reporting rate
attributes this drop to decreased reporting rate~(\cref{fig:TN_SIR_rep}),
whereas the SIR model with constant reporting rate attributes the drop
to decreased transmission rate~(\cref{fig:TN_beta_const}).
There is no evidence to corroborate which of the two interpretations
is correct.
Nevertheless, the SIR model with time-varying reporting rate
has a wider range of potential options, and it
explains the drop in cases via reduced reporting rate in April,
which is an explanation not available via the SIR model with constant reporting rate.

\Cref{fig:SIR_reporting} presents the estimated reporting rates
for the SIR model with time-varying reporting rate (in blue)
and with constant reporting rate (in orange).
In the case of New York~(\cref{fig:NY_SIR_rep}),
the $75\%$ and $95\%$ credible intervals
for the estimated constant reporting rate
are wider than the respective credible intervals
for the estimated time-varying reporting rate.
This indicates that our more flexible SIR model
reduces the uncertainty in reporting rate estimation
by letting the rate vary with time.
The time-varying reporting rate estimates
capture an upward trend in reporting rate
both in New York~(\cref{fig:NY_SIR_rep})
and in Tennessee~(\cref{fig:TN_SIR_rep}),
which can be explained by improvements in
infrastructure and in available resources
to manage the pandemic as time goes by.
On the other hand,
the SIR model with constant reporting rate
can not accommodate such temporal changes
in the management and reporting of the pandemic.

As seen in
~\cref{fig:SIR_reporting},
SIR modelling with constant reporting rate tends to underestimate reporting rates.
The disagreement in reporting rate estimation
between the SIR models with time-varying and with constant reporting rate
is particularly pronounced in the case of New York~(\cref{fig:TN_SIR_rep});
notice that the straight orange line (constant reporting rate estimate)
is lower than the blue line (time-varying reporting rate estimate).
The demonstrated underestimation of reporting rates 
via SIR modelling with constant reporting rate
has been previously
noted in the literature,
and it has been linked to underestimation of the true number of cases
and to bias in transmission rate estimation~\cite{gamado2014modelling}.
Our approach based on SIR modelling with time-varying reporting rate
provides a principled approach to avoid reporting underestimation,
consequently reducing the uncertainty in predictions of number of cases.

\subsection{Predictive uncertainty}

 \begin{figure}
	\centering
	\subfloat[SIR Predictions]
	{
		\includegraphics[width=0.47\textwidth]{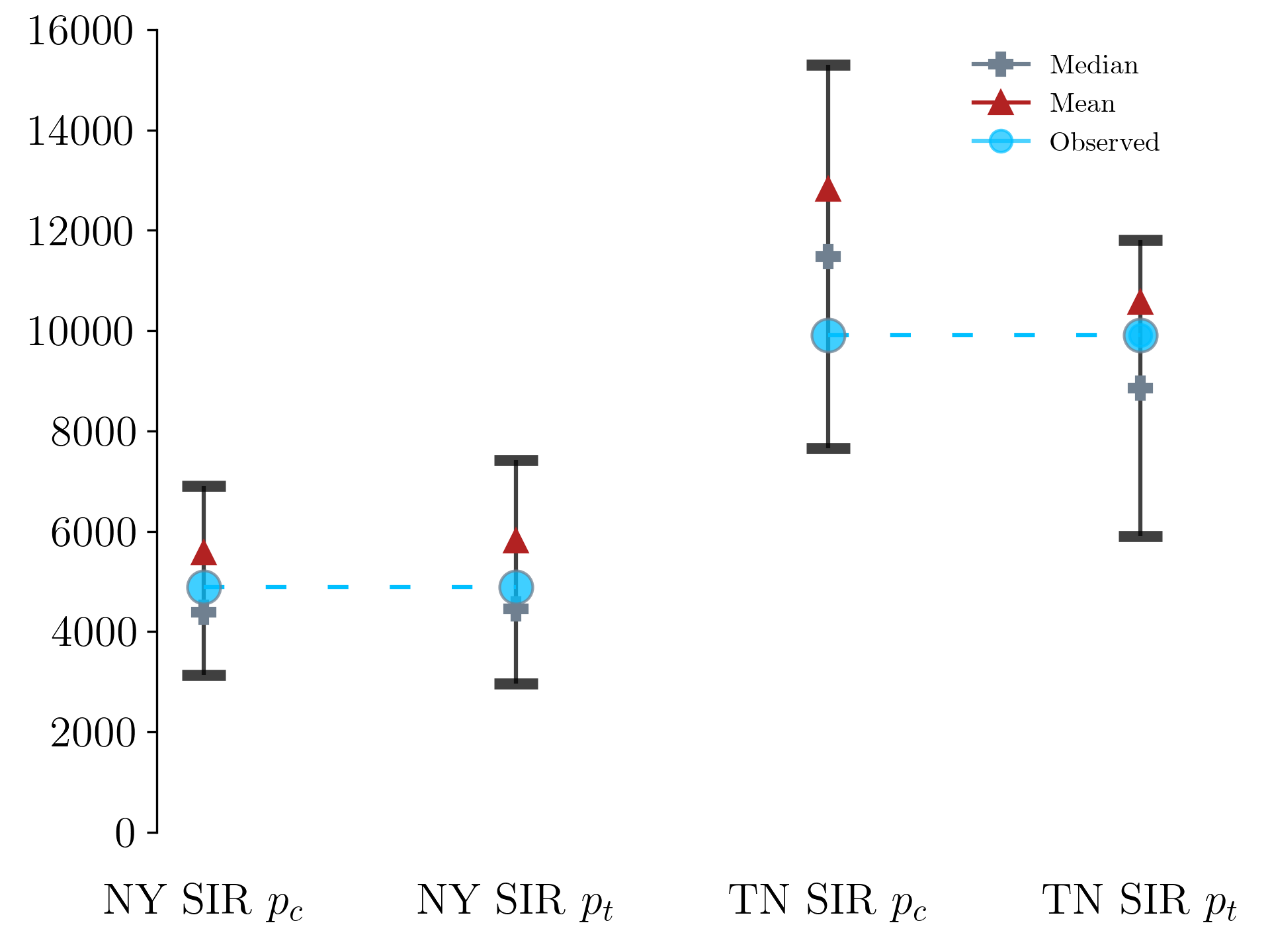}
		\label{fig:SIR_preds}
	}
	\hfill%
	\subfloat[SEIR Predictions]{
		\includegraphics[width=0.47\textwidth]{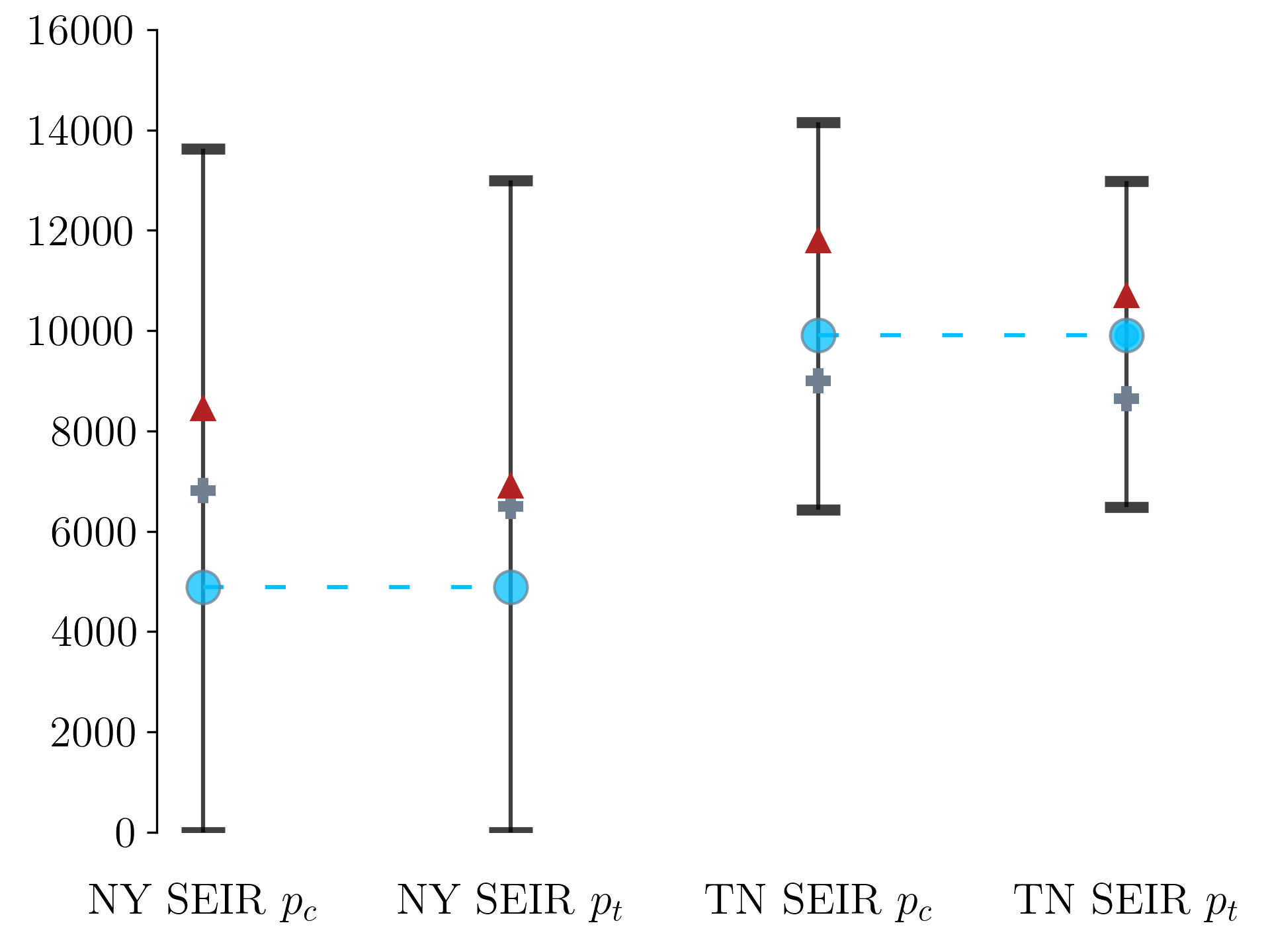}
		\label{fig:SEIR_preds}
	}
	\caption{One-week ahead predictions of COVID-19 cases
	in New York (NY) and in Tennessee (TN) for the week starting
	August 31, 2020,
	using SIR and SEIR models
	with constant reporting rate ($p_c$) or with time-varying reporting rate ($p_t$).
	Red triangles and grey crosses denote posterior predictive means and posterior predictive medians,
	respectively.
	Bars represent $75\%$ posterior predictive intervals.
	Blue circles depict the observed number of cases.}
	\label{fig:prediction_plots}
\end{figure}

We make predictions about the number of cases one week into the future
by fitting SIR and SEIR models with constant or with time-varying reporting rates
to New York and to Tennessee data through August 30, 2020.
Irrespective of whether constant or time-varying reporting rate is employed,
both SIR and SEIR models produce $75\%$ predictive intervals for
the number of cases in New York and in Tennessee that
contain the observed number of cases~(\cref{fig:prediction_plots}).

Overall,
SIR and SEIR models with time-varying reporting rates
outperform their counterparts with constant reporting rates
in terms of predictive performance~(\cref{fig:prediction_plots}).
Firstly, time-varying reporting rates yield narrower predictive intervals,
thus reducing predictive uncertainty.
Secondly, time-varying reporting rates lead to predictive posterior means
that are closer to the observed number of cases.
The improved predictive performance attained via modelling
based on time-varying reporting rates
is observed in three out of the four examined scenarios
(SIR and SEIR models fitted to Tennessee data,
and SEIR model fitted to New York data),
with no apparent improvement in one case (SIR model fitted to New York data).

The wider predictive intervals
and therefore higher predictive uncertainty in the Tennessee
predictions in comparison to the New York predictions shown in
~\cref{fig:prediction_plots} is seemingly counter-intuitive,
given the higher volatility of the New York data~(\cref{fig:data}).
However, there is an explanation for the higher predictive uncertainty associated with
the Tennessee data.
The Tennessee Department of Health changed how they defined an active case on September 3, 2020,
resulting in a one-day decrease of approximately $20{,}000$ reported cases 
~\cite{TN-health-dept}. Consequently, the predictive
distribution is skewed to the left in comparison to the
ground truth value, as anticipated after considering the change in the definition of
an active case by the Tennessee Department of Health.


\section{Discussion}\label{sect:conclusions} 

The states of New York and Tennessee experienced the first 
wave of the COVID-19 pandemic in different fashion. Our modelling strategy is able to dynamically
adapt to different mitigation strategies enacted in each locality and accurately 
reflect the course of the pandemic in these geographic regions. 
We are able to capture
 the dynamic nature of the transmission rate when intervention methods are
 enacted, and can quantify changes in the reporting rate of case counts.
 This modelling strategy yields actionable results for public health officials
entrusted with a community's well-being.


We observe dependence between the time-varying parameters, 
namely between the transmission and reporting rates, similar to the effects noted
by~\cite{gamado2014modelling}.
Indeed, with the significant under-reporting of active 
cases present in the New York data, a 
model employing a static reporting rate fails to capture the dynamic nature 
of COVID-19 transmissibility.
As a concrete example, consider the time period from March 1, 2020 through May 24, 2020, when there were $383{,}560$ active cases reported in New York. Taking the reporting rate inferred by our SIR model, we find
that there were $530{,}411$ active cases, with  95\%
confidence intervals $(403{,}285,\, 900{,}077)$, a figure which is corroborated
by the study of~\cite{albani2021covid} that identified under-reporting 
of active COVID-19 cases by considering hospitalization and death rates.


Primarily, a novelty in our modelling approach has been to include a time-varying reporting rate
that leads to models which are
more likely to fit and explain COVID-19 incidence data.
This conclusion is intuitive, since changes in the reporting rate
imply changes in the resulting data,
so a model with a varying reporting rate is more likely to fit
data affected by changes in reporting procedures.

Secondly, we provide a Bayesian approach to quantify uncertainty
in relevant epidemiological parameters and in predictions, yielding a source of
important information
to public health officials tasked with assessing the present state and
with suggesting mitigation
strategies for subsequent weeks.
Our one-week ahead predictions are accurate,
since $75\%$ relevant credible intervals contain the ground truth (\cref{fig:SIR_preds}).

The methods
described herein are better able to capture not only the time-varying
drivers of an epidemic, but also how the reporting of cases
changes temporally, thus providing more accurate quantification
of the spread of a disease through a susceptible population.
Our method provides near real-time actionable information to public health
officials, as opposed to methods that use the hospitalization rate~\cite{albani2021covid}
or the excess death rate~\cite{adam2022pandemic,lau2021evaluating}, both of which have a time-lag on the order of weeks. Indeed, previous studies have noted the presence of COVID-19
in February 2020, well before any appreciable increase in hospitalizations~\cite{covid2020evidence,subramanian2021quantifying}.
Quantifying the spread of a disease through a population
and the proportion that are going uncounted by public health agencies is an essential tool 
for these agencies tasked not only with estimating the proportion of a group that is
actively infected, but mitigating the disease's impact on a population.
Indeed, by providing real-time knowledge of the true number of active infections to
public health officials, the timing and severity of mitigation strategies can be 
better informed, thus reducing the community spread of a disease.

While our model cannot capture all the intricacies involved 
with the public health infrastructure, such as variability of testing sensitivity,
access to testing sites, or individuals taking at-home tests that are not
reported to public health agencies, we are able to estimate time-sensitive parameters crucial to slowing the spread of an emerging new disease. Indeed, by providing accurate and actionable information about the spread of a disease throughout a population, public health officials
could put in place mitigation strategies to slow the spread of a disease.

Future versions of the model could incorporate additional 
parameters, such as one describing mobility of subpopulations within a geographic region. 
Such a parameter could capture heterogeneity within a population, and identify those
subgroups at higher or lower risk for infection and transmission due to their 
movements within a specified time window.
Lastly, we plan to further investigate the correlation structure
between the transmission and reporting rates, to better quantify their 
dependencies and effects on each other.


\section{Methods}\label{sect:modelling}

\subsection{Bayesian formulation}\label{sect:Bayes}

Our SIR and SEIR models are parameterized by
$\theta = (\theta_w, \theta_X, \theta_Y)$. 
We factorize
the posterior density
$\pi(w_{0:T}, p_{0:T},\theta\given Y_{1:T})$
of the transmissibility $w_{0:T}$,
reporting rate $p_{0:T}$ 
and model parameters $\theta$, 
given observations $Y_{1:T}$,
as follows:
\begin{equation}
	\pi(w_{0:T}, p_{0:T},\theta\given Y_{1:T}) = \pi(w_{0:T},p_{0:T}\given Y_{1:T},\theta)\pi(\theta\given Y_{1:T}).
	\label{eq:bayes_decomp}
\end{equation}
According to~\cref{eq:bayes_decomp},
we sample from $\pi(w_{0:T}, p_{0:T},\theta\given Y_{1:T})$
by alternating between sampling from densities 
$\pi(w_{0:T},p_{0:T}\given Y_{1:T},\theta)$ and 
$\pi(\theta\given \Ytraj)$
via the particle Markov chain Monte Carlo
(PMCMC) algorithm of~\cite{andrieu2010particle}.
PMCMC
alleviates issues of convergence and insufficient exploration of
the sample space that can arise due to correlations
and dependencies between variables. 

Sampling from \cref{eq:bayes_decomp}
allows us to infer the time-varying transmission rate $\beta_{0:T}$,
the time-varying reporting rate $p_{0:T}$,
and to make predictions about the future course of
the pandemic. Moreover, our Bayesian SIR and SEIR models
enable us to quantify the uncertainty of our parameter estimates and of our predictions.

\subsection{Particle Markov chain Monte Carlo}\label{sect:MCMC}

To sample from the posterior density of~\cref{eq:bayes_decomp},
	we employ PMCMC sampling~\cite{andrieu2010particle}. We describe the algorithmic procedure
	and detail the hyperparameter choices in our model; for an
	in-depth discussion and theoretical results,
	see~\cite{andrieu2010particle,del2006sequential}. 
 	PMCMC alleviates issues with slow MCMC mixing and low acceptance rates that 
 	are present in other methodologies for sampling from a joint posterior, such as the 
 	pseudo-marginal approach of~\cite{andrieu2009pseudo}. First, the
	sequential Monte Carlo (SMC) procedure is described, followed by PMCMC.
	
	SMC algorithms~\cite{del2006sequential} provide a way of
	sampling from distributions defined by
	state-space models. 
	Based on the decomposition of our posterior density as stated in~\cref{eq:bayes_decomp},  
	samples are first drawn from the conditional density
	$\pi(w_{0:T}, p_{0:T} \given Y_{1:T}, \theta)$.
	The employed SMC algorithm
	yields a sequence of densities that approximates 
	$\{\pi(w_{0:\tau}, p_{0:\tau}\given Y_{1:\tau},\theta):\tau\geq0\}$
	and the marginal densities
 	$\{\cL(Y_{1:\tau}\given w_{0:\tau}, p_{0:\tau}, \theta):\tau\geq0\}$
 	for a given $\theta$ and
 	$\tau\leq T$.
 	SMC
 	first approximates $\pi(w_1, p_1 \given Y_1,\theta)$ and
 	$\cL(Y_1\given w_{0:1}, p_{0:1}, \theta)$ by drawing samples from an 
 	importance density
 	$\hat{q}^{(i)}_1\sim Q_1(\cdot\given Y_1, w_1, p_1, \theta)$ for 
 	particles $i=1,\dots,P$ and 
 	$q^{(i)}_t := (w^{(i)}_t, p^{(i)}_t)$~\cite{andrieu2010particle}. 
 	SMC
 	approximates $\pi(q_{1:\tau} \given Y_{1:\tau},\theta)$ 
 	and $\cL(Y_{1:\tau}\given q_{0:\tau},\theta)$,
 	for subsequent iterations $\tau$ by sampling from importance densities
 	$\hat{q}^{(i)}_\tau\sim Q_\tau(q_\tau \given Y_{1:\tau}, \hat{q}^{(i)}_{1:\tau-1}, \theta)$.
 	Requiring these densities to be of the form 
 	$Q_\tau(q_{1:\tau}) = Q_{1}(q_{1})\prod_{\tau=2}^T\,Q_\tau(q_\tau \given Y_{1:\tau},  q_{1:\tau-1})$, 
 	one readily computes an unbiased estimate of the marginal likelihood $\cL(Y_{1:T}\given q_{0:T},\theta)$,
 	which is necessary for 
 	the Metropolis-Hastings acceptance ratio in the PMCMC sampler~\cite{del2006sequential}.

 	Having sampled from $\pi(w_{0:T}, p_{0:T}\given Y_{1:T},\theta)$ via SMC, 
 	it remains to sample from $\pi(\theta\given\Ytraj)$.  
 	At each iteration of the PMCMC algorithm, a value $\theta^*$ of the parameter $\theta$ is proposed,
 	followed by a sample $\{q_{0:T}^{(i)}\}_{i=1}^P$ generated via SMC.
 	Thus, the problem of sampling from
 	$\pi(w_{1:T}, p_{1:T}, \theta\given Y_{1:T})$ is reduced to sampling from $\pi(\theta\given Y_{1:T})$,
 	as samples from $\pi(w_{1:T}, p_{1:T} \given Y_{1:T}, \theta)$ are obtained via the SMC algorithm.

The model parameters $\eta, w_0, \sigma$, and $p_c$ or $p_t$
are given wide uninformative priors
due to the uncertainty about the ongoing pandemic and disparities in reporting
data. We model the infection period as a truncated Gaussian distribution
with mean of $5.058$ days, standard deviation of $1.51$,
lower bound of $2.228$ days and upper bound of $11.8$ days,
following~\cite{lauer2020incubation}. 
The prior for the latent period $E_t$ is obtained from the study of~\cite{moghadas2020implications}, and
is modeled as a gamma distribution with shape and scale parameters
$1.058$ and $2.174$
respectively~\cite{he2020temporal}. For the initial proportions of the population
in states $X_0$ we chose a Dirichlet distribution, while constraining the mean of $R_0$
to be $\cN(0.5, 0.25^2)$, and let the means of the other compartments be equal. By this 
choice, we ensure that the condition $S_t + E_t + I_t + R_t = N$ or  $S_t + I_t + R_t = N$
is satisfied in the respective SEIR or SIR model. Thus, the
sum over all compartments in the epidemiological model at each time step 
is the same as the total population $N$. Lastly, we ran
PMCMC sampling
with $5,000$ particles and obtain $50,000$ samples from the posterior after
a burn-in period of $5,000$ iterations.

\subsection{Choice of density for the observational model}

A Poisson or a Gaussian approximation can be used for
the density $h(Y_{1:T}\given X_{0:T}, p_{0:T}, \theta_{Y})$ of the observational model.
Pilot PMCMC runs demonstrate similar effective sample sizes
for the Poisson and Gaussian approximations,
but higher number of particles and therefore higher computational budget
are required for the Poisson approximation.
For this reason, a Gaussian approximation is preferred.

\subsection{Overview of data and of experimental setup}    

The data used in our experiments are based on
daily case counts
from March 1, 2020, through August 31, 2020,
obtained from the New York Times COVID data 
repository~\cite{nytimes-covid}.
In our analysis, we use daily reported case counts and 
aggregate them on a weekly basis for computational considerations. 
For one iteration of the PMCMC method,
each particle in the ensemble requires
the numerical approximation of a system of non-linear
ordinary differential equations comprised of $T$ time steps.
This computational cost becomes infeasible in the case of daily case counts
due to the increased number of particles required for PMCMC sampling.

For the implementation of our model and for PMCMC sampling,
we use the Bayesian modelling software \texttt{libBi}~\cite{murray2013bayesian}
and the R packages \texttt{rbi} and \texttt{rbi.helpers}~\cite{rbi-helpers,rbi}.
Our models, data and code for reproducing our results can be found at
\url{https://github.com/aspannaus/Covid-model}.

\section*{Acknowledgments}

This manuscript has been authored by UT-Battelle, LLC under Contract No. DE-AC05-00OR22725 
with the U.S. Department of Energy. The United States Government retains and the publisher, 
by accepting the article for publication, acknowledges that the United States Government 
retains a non-exclusive, paid-up, irrevocable,world-wide license to publish or reproduce 
the published form of this manuscript, or allow others to do so, for United States 
Government purposes. The Department of Energy will provide public access to these 
results of federally sponsored research in accordance with the DOE Public 
Access Plan (http://energy.gov/downloads/doe-public-access-plan).

Research was supported by the National 
Virtual Biotechnology Laboratory, 
a consortium of DOE national laboratories focused on response to 
COVID-19, with funding provided by the Coronavirus CARES Act.
All numerical experiments were completed employing 
the computing resources of the Compute and Data Environment for Science (CADES) 
at the Oak Ridge National Laboratory. The latter of which is supported by the 
Office of Science of the U.S. Department of Energy under
Contract No. DE-AC05-00OR22725. 

\section*{Author contributions statement}

A.S. and T.P. conceived the experiments and analysed the results. 
A.S. developed the models, conducted the experiments, created the figures, 
and wrote the initial draft of the manuscript.
A.S. and T.P. edited and finalized the draft.
All authors reviewed and commented on the manuscript. 
J.B.C and S.E. secured initial funding and project conceptualization.

\section*{Competing interests}

The authors declare no competing interests.




\newpage
\appendix

\begin{appendices}

\section{SEIR model definition}\label{seir_model_def}

The SEIR epidemiological model with time-varying transmission rate
$\beta (t)$
is represented by the
system of differential equations
	\begin{equation*}
	    \begin{aligned}	
	    \der S(t) &= -\beta(t)S(t)\frac{I(t)}{N} \der t, \\
	    \der E(t) &= (\beta(t)S(t)\frac{I(t)}{N} - \kappa E(t)) \der t, \\
	    \der I(t) &= (\kappa E(t) - \gamma I(t))	\der t, \\
	    \der R(t) &= \gamma I(t)\, \der t,\\
	    \der w(t) &= \mu(w(t), \theta_w)\der t + \sigma(w(t),\theta_w)\,\der B(t), \quad
	        w(t) = g(\beta(t)).\label{eq:SEIR}
	    \end{aligned}
	\end{equation*}
	In this study, we take $\mu=0$ and $\theta_w=\sigma$, thus assuming that the transmission rate 
	on any day is likely to be the same as the previous day, and set $g(\cdot) = \log(\cdot)$.
	By choosing $\mu=0$ and $\theta_w=\sigma$, the resulting path 
	$w(t)$, defined by the stochastic differential equation
	$\der w(t)=\theta_w\der B(t)$, is Brownian motion.
	Allowing $w(t)$ to vary in time defines
	the effective contact rate $\beta(t)$, which controls the extent to which
	a disease spreads between individuals within a population.

\section{SEIR model results}\label{si:poisson_results}

\begin{figure}
	\centering
	\subfloat[]
	{
			\label{fig:NY_SEIR_beta_const}
		\includegraphics[width=0.46\textwidth]{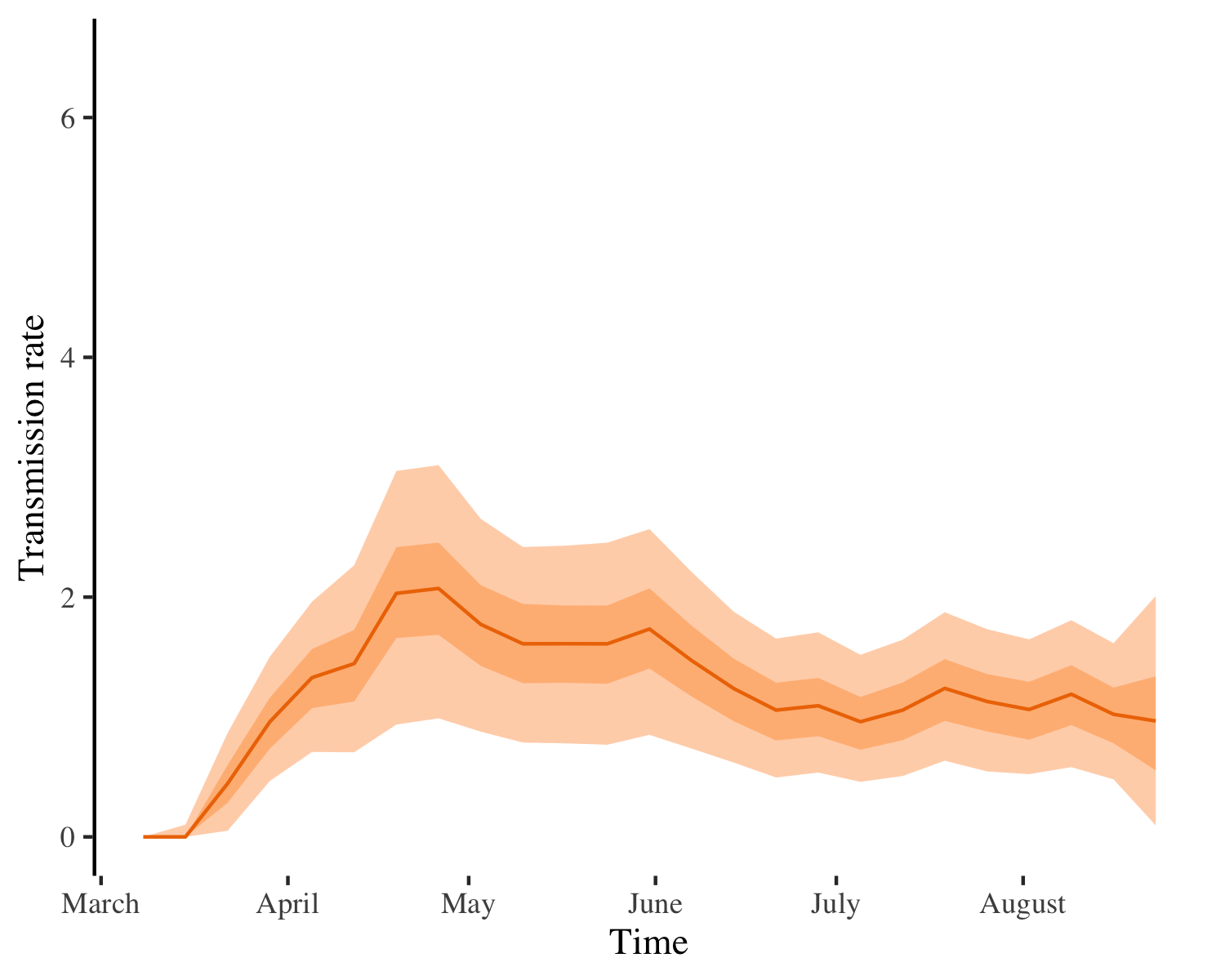}
	}
	\hfill
	\subfloat[]
	{
			\label{fig:NY_SEIR_beta_SS}
		\includegraphics[width=0.46\textwidth]{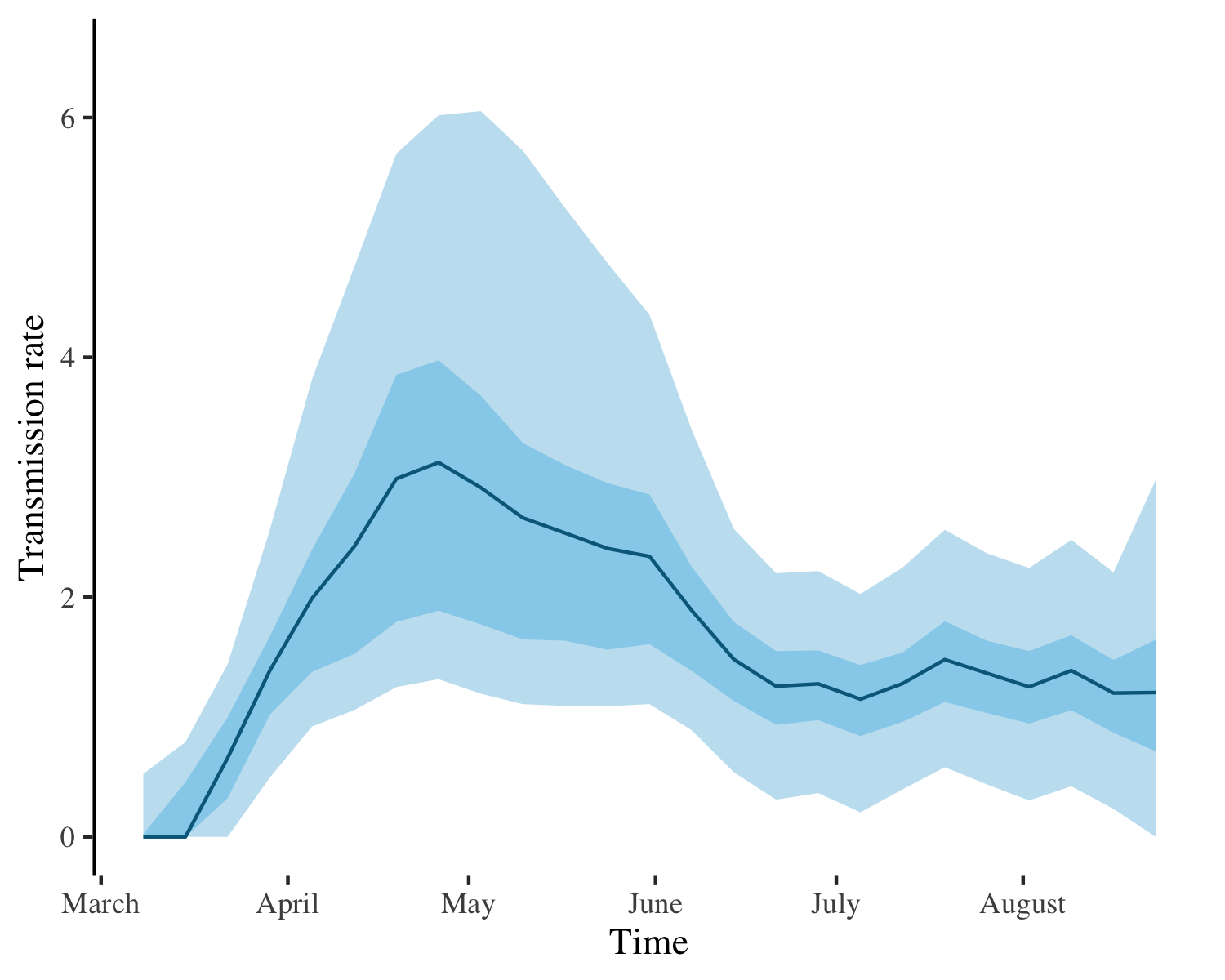}
	}
	\caption{Estimates of the COVID-19 transmission rate for New York
		based on the typical SEIR model with constant reporting rate~(\cref{fig:NY_SEIR_beta_const}) and
		based on our SEIR model with a time-varying reporting rate $p_t$~(\cref{fig:NY_SEIR_beta_SS}).
		Solid lines, light-shaded and dark-shaded areas
		correspond to posterior means, $75\%$ and $95\%$ credible intervals
		of the associated transmission rates.}\label{fig:NY_SEIR_beta}
\end{figure}

\begin{figure}
	\centering
 	\subfloat[]
	{
			\label{fig:TN_SEIR_beta_const}
		\includegraphics[width=0.46\textwidth]{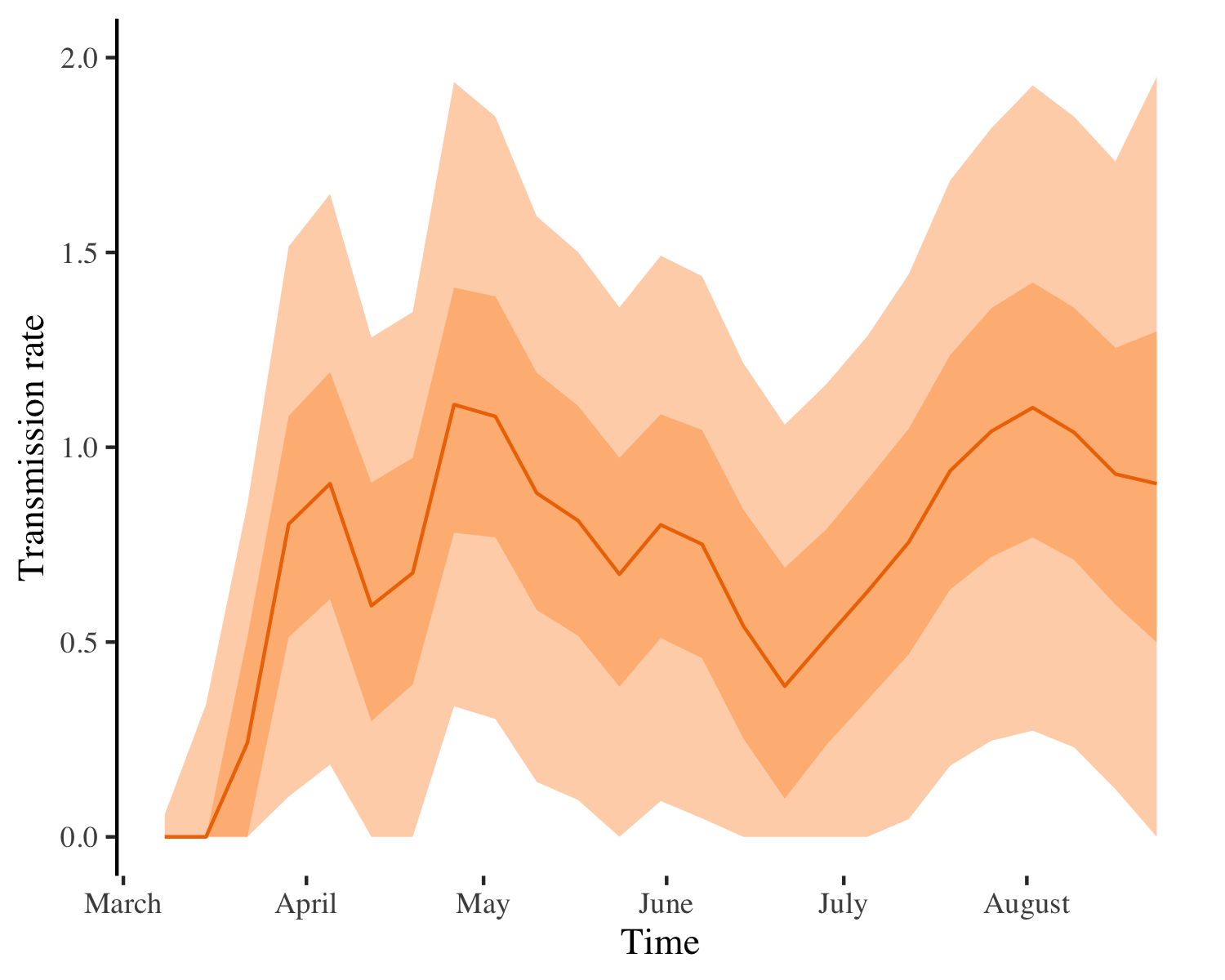}
	}
	\hfill
 	\subfloat[]
	{
		\label{fig:TN_SEIR_beta_SS}
		\includegraphics[width=0.46\textwidth]{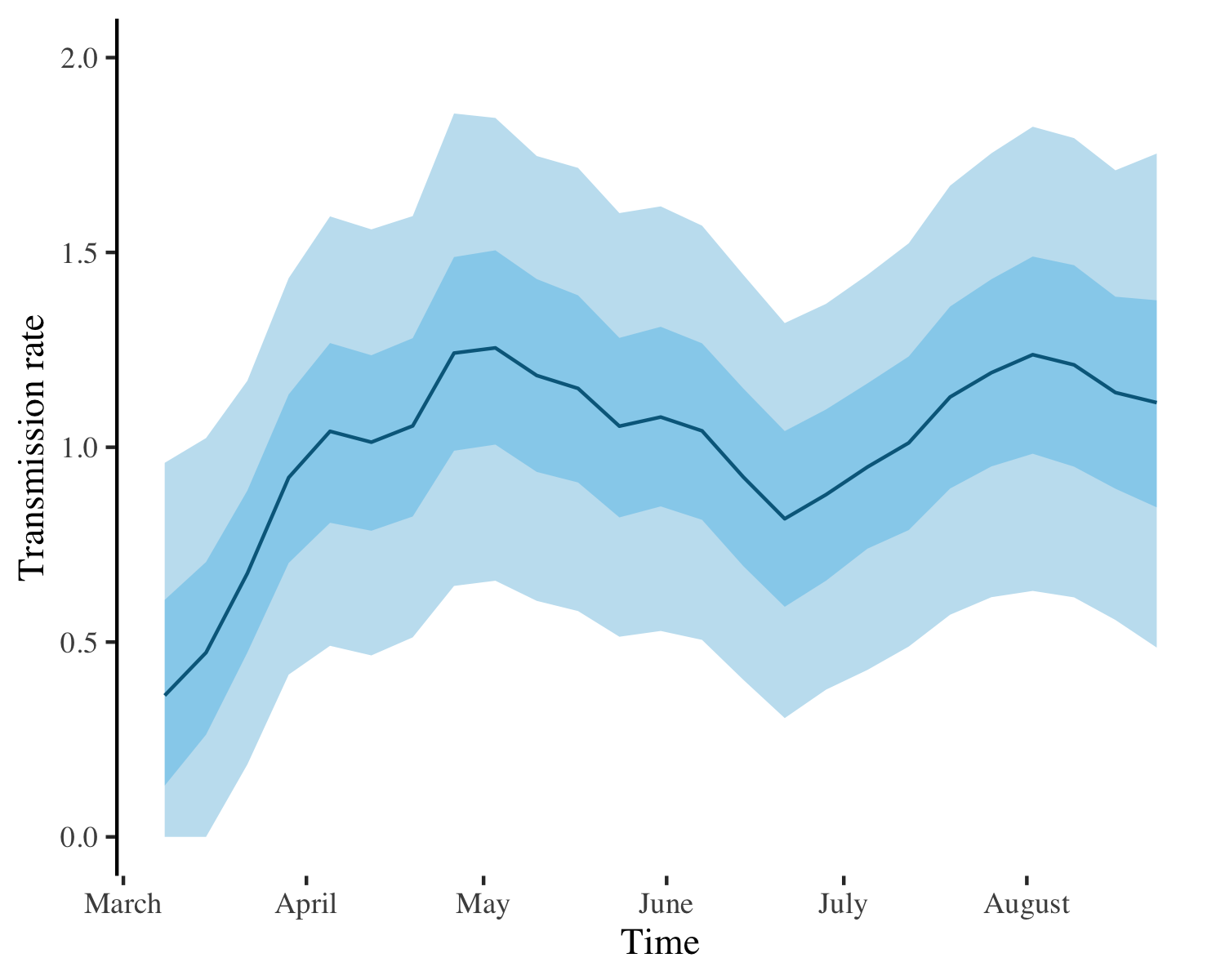}
	}
	\caption{Estimates of the COVID-19 transmission rate for Tennessee
		based on the typical SEIR model with constant reporting rate~(\cref{fig:TN_SEIR_beta_const}) and
		based on our SEIR model with a time-varying reporting rate $p_t$~(\cref{fig:TN_SEIR_beta_SS}).
		Solid lines, light-shaded and dark-shaded areas
		correspond to posterior means, $75\%$ and $95\%$ credible intervals
		of the associated transmission rates.}\label{fig:SEIR_TN_beta}
\end{figure}

\begin{figure}
	\centering
	\subfloat[New York]
	{
			\label{fig:NY_reporting}
		\includegraphics[width=0.46\textwidth]{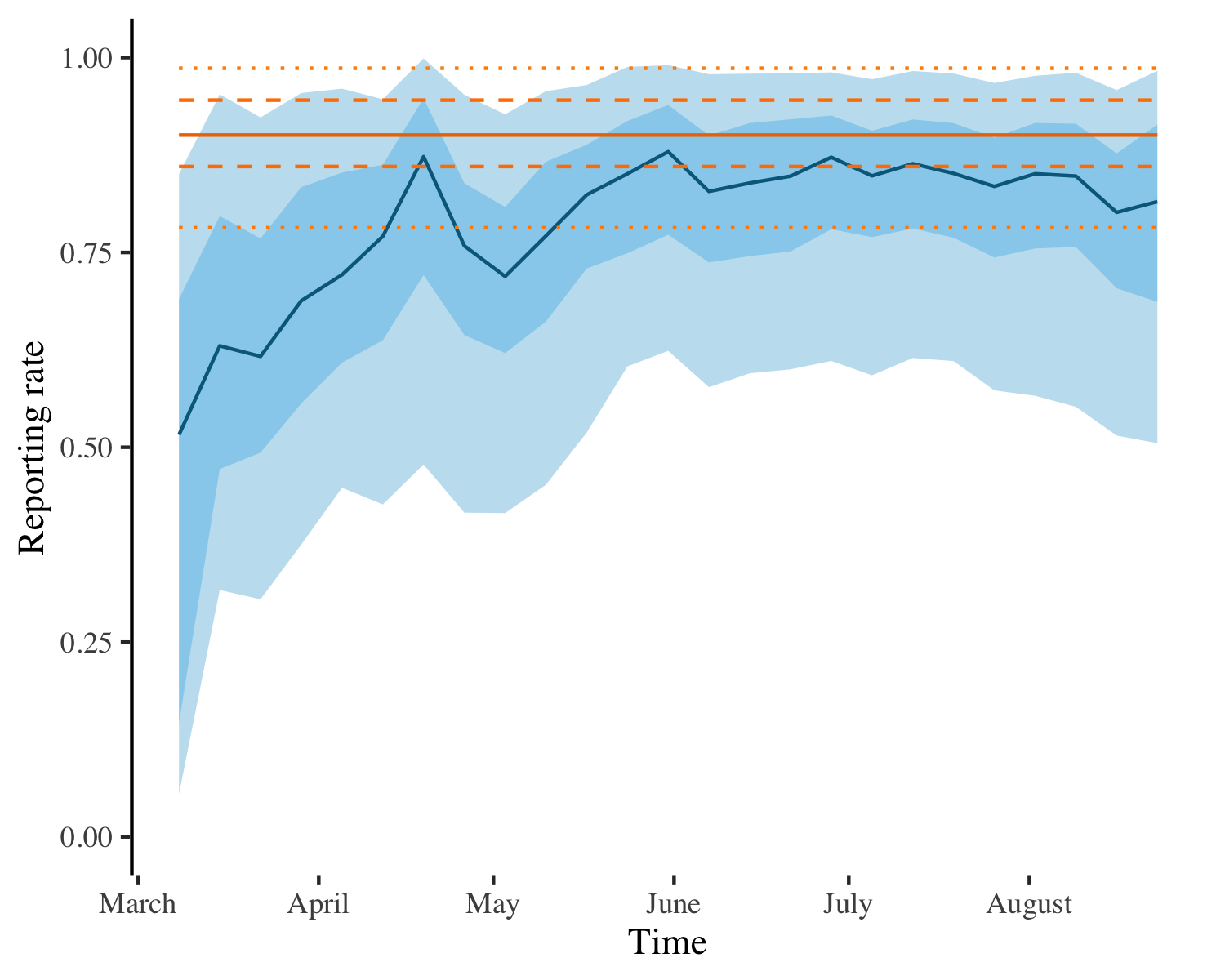}
	}
	\hfill
	\subfloat[Tennessee]
	{
			\label{fig:TN_reporting}
		\includegraphics[width=0.46\textwidth]{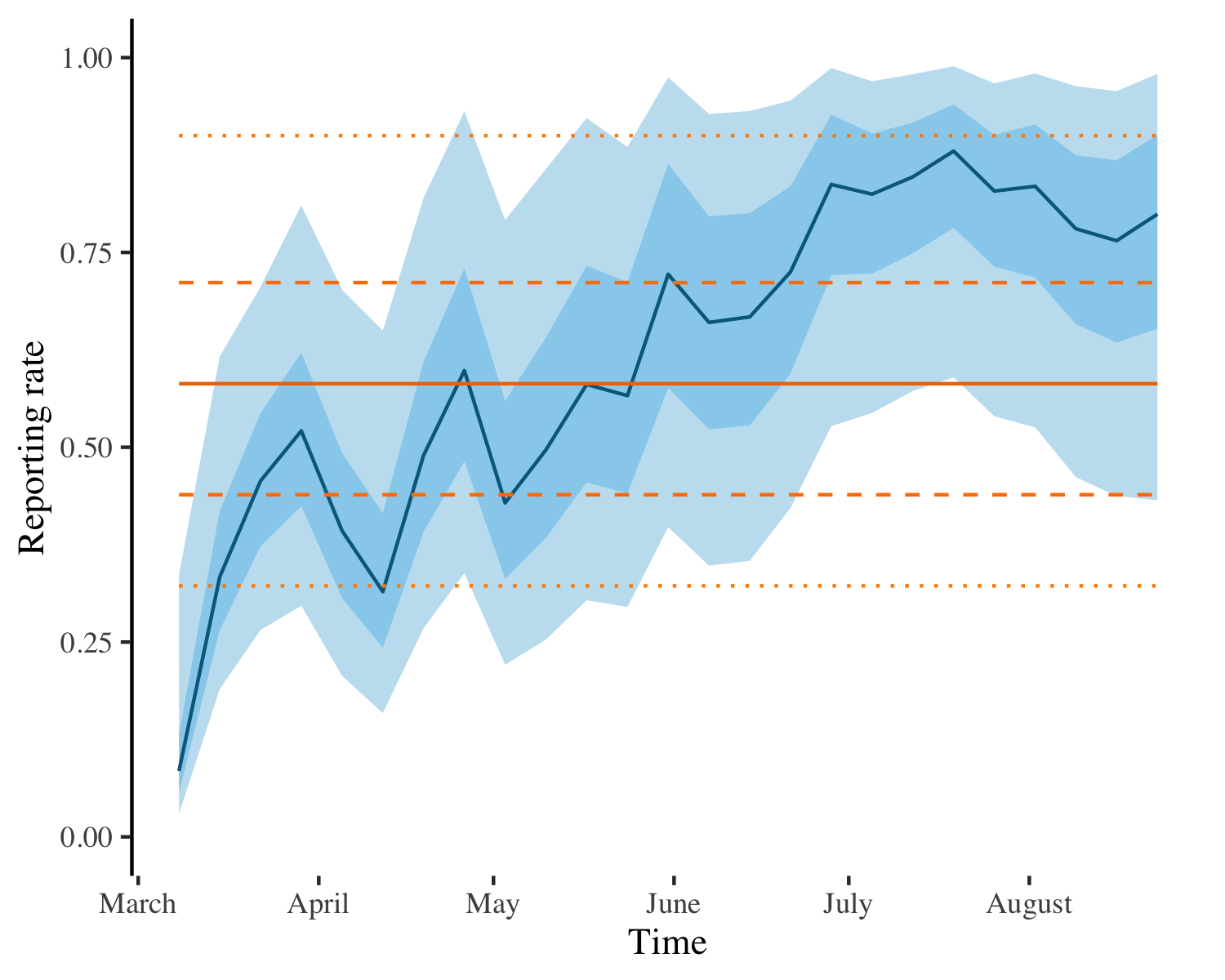}
	}
	\caption{Estimates of the COVID-19 reporting rate
	for New York~(\cref{fig:NY_reporting}) and for Tennessee~(\cref{fig:TN_reporting})
	obtained from the SEIR model with constant (orange) and with time-varying (blue)
    reporting rate. 
	The blue line, light-shaded and dark-shaded blue areas
	represent the respective posterior mean,
	$75\%$ and $95\%$ credible intervals of the time-varying reporting rate.
	The orange solid, orange dashed and orange dotted line
	represent the respective posterior mean,
	$75\%$ and $95\%$ credible intervals of the constant reporting rate.}\label{fig:SEIR_reporting}
\end{figure}

In this appendix, we present results on parameter uncertainty
based on SEIR modelling,
while the main manuscript presents results on parameter uncertainty
based on SIR modelling.
We observe
that the inclusion of the latent infected compartment in the epidemiological model has a marked impact on the estimate of the time-varying reporting and transmission rates
~(\cref{fig:NY_SEIR_beta}, \cref{fig:SEIR_TN_beta} and \cref{fig:SEIR_reporting}).
The effect on the latter follows from recalling that the transmissibility is the product of the probability of passing the disease to another individual and the number of interactions with all 
individuals. If there is in fact a latent phase in the course of COVID-19,
then it follows that individuals in the exposed bin could be mixing with the
general population, potentially passing on the disease as it transitions from a latent to active infection.

In the case of New York data,
the posterior mean estimate of the constant reporting rate of the SEIR model
has narrower credible intervals~(straight orange line in \cref{fig:NY_reporting})
than the posterior mean estimate
of the constant reporting rate of the SIR model~(\cref{fig:NY_SIR_rep}).
However, the credible intervals for the former seem spurious,
since the the posterior mean estimate of the time-varying reporting rate of our SEIR model
varies substantially with time~(blue line in \cref{fig:NY_reporting}).
In fact, the posterior predictive means of number of cases are closer to the
respective observed number of cases,
when employing a time-varying (rather than constant) reporting rate in the SEIR model
~(\cref{fig:SEIR_preds}).
So, the reduced predictive capacity of the SEIR model with constant reporting rate
(in comparison to our SEIR model with time-varying reporting rate)
implies further that the estimated constant reporting rate in
\cref{fig:NY_reporting} is not accurate.


\end{appendices}

\end{document}